\documentclass{article}

\PassOptionsToPackage{numbers,sort}{natbib}
\usepackage[main,final]{neurips_2026}
\setcitestyle{numbers,square,comma}
\renewcommand{\acksection}{\section*{Acknowledgments}}

\usepackage[utf8]{inputenc}
\usepackage[T1]{fontenc}
\usepackage{float}
\usepackage{hyperref}
\hypersetup{hidelinks}
\usepackage{url}
\usepackage{booktabs}
\usepackage{amsfonts}
\usepackage{amsmath}
\usepackage{amssymb}
\usepackage{nicefrac}
\usepackage{microtype}
\usepackage{xcolor}
\usepackage{graphicx}
\graphicspath{{figures/}}
\usepackage{wrapfig}
\usepackage{multirow}
\usepackage{enumitem}
\usepackage{xspace}
\usepackage{makecell}
\usepackage{placeins}
\usepackage{needspace}
\newcommand{\honset}{\ensuremath{\mathbf{h}_{\text{onset}}}\xspace}
\newcommand{\hpre}{\ensuremath{\mathbf{h}_{\text{pre}}}\xspace}
\newcommand{\pdbs}{\textsc{PDBS}\xspace}
\newcommand{\cbl}{\textsc{CBL}\xspace}
\newcommand{\Evt}{\ensuremath{E_{vt}}\xspace}
\newcommand{\Eat}{\ensuremath{E_{at}}\xspace}
\newcommand{\Hone}{\ensuremath{H_{1}}\xspace}
\newcommand{\jf}{\ensuremath{\mathcal{J}\&\mathcal{F}}\xspace}
\newcommand{\delt}{\ensuremath{\delta}\xspace}

\title{To Think or Not to Think: Pre-Decisional Reasoning Budgets for Referring Audio-Visual Segmentation}

\author{%
  Haojie Hu\textsuperscript{2}\thanks{Equal contribution.}\quad
  Senda Chen\textsuperscript{2}\footnotemark[1]\quad
  Ying Shen\textsuperscript{2}\thanks{Corresponding author.}\quad
  Lin Zhang\textsuperscript{1,2} \\
  \textsuperscript{1}IEIT SYSTEMS (Beijing) Co., Ltd. \\
  \textsuperscript{2}School of Computer Science and Technology, Tongji University, Shanghai, China \\
  \texttt{\{2351493,2411498,yingshen,cslinzhang\}@tongji.edu.cn}
}

\begin{document}

\maketitle

\begin{abstract}
Multimodal reasoning systems increasingly assume that longer chain-of-thought (CoT) uniformly improves downstream grounding. We show that this assumption fails in referring audio-visual segmentation (Ref-AVS): simple queries can be harmed by excessive reasoning, creating an ``overthinking trap'' associated with phrase and localization drift, while genuinely ambiguous queries still benefit from multi-step reasoning. To systematically study this, we construct counterfactual reasoning-budget labels under a five-fold out-of-fold evaluation protocol. Our results recast adaptive Ref-AVS reasoning as a pre-decisional representation problem: before the first generated token, the generation-onset state contains linearly decodable information predictive of whether additional reasoning will improve segmentation. A linear probe predicts binary reasoning need with 70.1\% accuracy, substantially outperforming text-only features and uncertainty-based proxy signals. As a router, the onset-state controller improves over downstream confidence routing and closely matches \emph{Short-then-Decide} while using far fewer tokens. Proxy comparisons show that the onset representation contains predictive information not fully recovered by the evaluated hand-crafted features. Reusing this onset state for lightweight, single forward-pass routing reduces autoregressive token consumption by 60\% while retaining about 96\% of always-long performance. It further provides a cost--quality trade-off on R$^2$-AVSBench without target-label tuning, rather than an accuracy improvement over always-long. Error analysis links harmful long reasoning to phrase and localization drift, illustrating the benefit of bypassing unnecessary reasoning on simple queries. Ultimately, our results support efficient adaptive reasoning in Ref-AVS through pre-decisional internal representations.
\end{abstract}

\section{Introduction}
\label{sec:intro}

\begin{figure}[t]
\centering
\includegraphics[width=\linewidth]{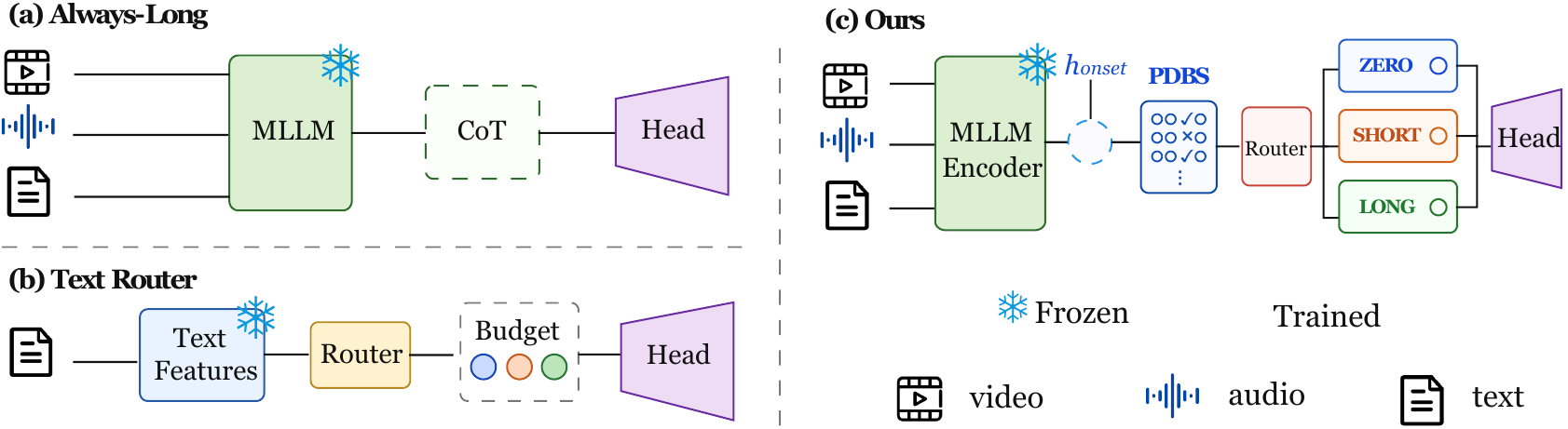}
\caption{\textbf{Overview of pre-decisional reasoning budgets for Ref-AVS.} (a) The conventional always-long pipeline sends every video--audio--text query through a frozen MLLM and a full CoT before grounding, implicitly assuming that maximal reasoning is universally beneficial. (b) A text-router baseline predicts the reasoning budget from frozen textual features alone, but cannot access the integrated multimodal state where sample-specific reasoning need emerges. (c) Our framework reuses the frozen Ref-AVS MLLM encoder to capture the generation-onset representation \honset before the first reasoning token. The resulting Pre-Decisional Budget Signal (\pdbs) is read by a lightweight router, which decides whether each sample should use \textsc{Zero}, \textsc{Short}, or \textsc{Long} reasoning before segmentation. Snowflakes denote frozen modules.}
\label{fig:teaser}
\end{figure}

Multimodal large language models (MLLMs) have rapidly advanced complex scene understanding by unifying reasoning across language, vision, and audio~\citep{vaswani2017attention, radford2021clip, alayrac2022flamingo, li2023blip2, liu2023llava}. In Ref-AVS---where the goal is to localize sounding objects in dynamic scenes from natural language queries~\citep{wang2024refavs}---state-of-the-art systems~\citep{tgs2026, aurora2025, mar3_2026} now insert CoT reasoning~\citep{wei2022cot, kojima2022zeroshot} before spatial grounding, using the generated language to refine cross-modal alignment on the premise that more reasoning always improves segmentation.

We show that this premise is wrong. For many queries, the visual evidence is already unambiguous: a speaker clearly visible against a wall, a single instrument in the frame. Forcing the model through a lengthy reasoning chain on such inputs can create what we call the \emph{overthinking trap}---the generated phrase adds unsupported details, drifts from the visual evidence, and ultimately produces a worse segmentation mask. Conversely, genuinely ambiguous references---such as identifying which of several instruments plays the longest---still benefit from extended multi-step reasoning. In other words, the value of reasoning is not a property of the model alone, but of the particular input-model interaction: the same reasoning machinery can clarify one scene while distracting from another. To quantify this, we run each sample under all candidate reasoning policies and compare the outcomes. The heterogeneity is pervasive: no fixed budget is preferred for every sample, and an idealized per-sample policy substantially outperforms the standard always-long strategy (Table~\ref{tab:heterogeneity}).

This leads to a fundamental question: \emph{can reasoning-budget utility be predicted from the model state before reasoning begins?} Rather than evaluating reasoning need from external cues---text-only features, post-hoc confidence scores, or staged generation pipelines---we look inside the model. Specifically, we examine the MLLM's internal hidden state at the moment just after it has processed all multimodal inputs but before it has generated the first reasoning token. We call this the \emph{generation-onset representation}, \honset. This state is pre-decisional: it has already integrated video, audio, and text evidence, yet contains no generated reasoning content. This framing treats budget allocation as supervised prediction from multimodally integrated states, without assuming that the model itself computes a judgment of reasoning need.

We find that \honset encodes an informative signal about reasoning need. Even a single-layer linear classifier trained on this state---without access to any generated text---can predict whether additional reasoning will help, significantly outperforming text-only and uncertainty-based baselines. To characterize \pdbs, we test whether it can be recovered from interpretable low-dimensional proxies designed to capture modality-level uncertainty and alignment. These comparisons indicate that the evaluated hand-crafted features do not fully recover the predictive information available in the onset state; they do not establish an irreducibly high-dimensional mechanism.

Building on this finding, we design a lightweight, single forward-pass controller that reads \honset and dynamically assigns each sample a reasoning budget---skip reasoning entirely, reason briefly, or reason at full depth---without any additional model forward pass. Because the onset state is already computed during input encoding, the controller adds less than 1\,ms after that encoding. The resulting system behaves less like a fixed-depth reasoner and more like a selective reasoner: it can preserve direct visual grounding when the reference is clear, while invoking deeper multimodal reasoning for disambiguation. It reduces reasoning token consumption while retaining nearly all of the performance of always-long reasoning; a related cost--quality trade-off is observed without target-label tuning on R$^2$-AVSBench.

\Needspace{14\baselineskip}
The main contributions are as follows:
\begin{itemize}[leftmargin=1.4em,itemsep=2pt,topsep=4pt,parsep=0pt]
    \item The \emph{overthinking trap} is characterized through matched reasoning-budget outcomes: forced CoT can degrade segmentation on simple inputs as well as waste computation. Formalizing budget heterogeneity reveals the advantage of a cost-aware oracle that selects the cheapest evaluated budget within a quality tolerance.
    \item We provide representation-level evidence that reasoning-budget utility is predictable \emph{before the first token is generated}. A linear classifier on the generation-onset state significantly outperforms text-only and uncertainty-based baselines; readout ablations and stronger proxy comparisons characterize the information available to an external budget controller.
    \item We translate this finding into a practical system: a single forward-pass controller that reads \honset and dynamically assigns each sample a reasoning budget, bypassing external difficulty evaluators entirely. This cuts token consumption by 60\% while preserving about 96\% of always-long \jf, with a cost--quality trade-off also observed on the evaluated R$^2$-AVSBench split.
\end{itemize}

\section{Related Work}
\label{sec:related}

\paragraph{Reasoning in Ref-AVS.}
Ref-AVS inherits the grounding difficulty of referring-expression, image, and video segmentation~\citep{kazemzadeh2014referit, hu2016segmentation, yang2022lavt, seo2020urvos, botach2022mttr, wu2022referformer, li2023r2vos}, while adding audio as a disambiguating modality~\citep{wang2024refavs}. Early and perception-centric pipelines emphasize stronger grounding and mask generation, including audio-queried fusion, open-vocabulary detection, promptable segmentation, temporal propagation, and instance-level extensions~\citep{li2023catr, li2022glip, liu2023dino, kirillov2023sam, ravi2024sam2, ren2024gdsam2, radman2025tsam, wang2025sam2love, liu2024ufe, huang2025vct, guo2025avis}. A complementary line studies counterfactual learning to mitigate spurious audio-visual correlations~\citep{zha2025icf}. Recent methods increasingly add heavy reasoning: TGS-Agent~\citep{tgs2026} introduces a structured Think-Ground-Segment chain; AURORA~\citep{aurora2025} uses reflective GRPO training; MAR3~\citep{mar3_2026} dispatches multi-agent reasoning based on reference difficulty; and OmniAVS~\citep{omniavs2025} expands modality coverage. Explicit reasoning and agent-based designs also appear in CoT-RVS, VideoSeg-R1, and Refer-Agent~\citep{kao2026cotrvs, xu2025videosegr1, jiang2026referagent}. More broadly, recent work studies certainty-based adaptive routing and CoT efficiency for LLMs/MLLMs~\citep{wang2025car, tang2025mmecot, xu2025adactrl, zhang2026cotseg}. Our focus is the combination of a fused pre-generation readout, counterfactual mask-utility supervision, and budget selection before the first reasoning token in Ref-AVS. MAR3~\citep{mar3_2026} contains an empirical hint of over-reasoning, where excessive reflection steps ($N{=}3$) underperform fewer steps ($N{=}2$); we study its sample-level consequences through matched budget outcomes.

\paragraph{Probing and Mechanistic Interpretability.}
Linear probing is widely used to reveal information encoded in neural representations~\citep{alain2017probing, conneau2018cram, tenney2019bert, belinkov2022probing}, while control tasks and selectivity analyses clarify when a probe reflects usable structure rather than memorization~\citep{hewitt2019control}. Mechanistic interpretability further uses circuit analysis, causal tracing, activation patching, and model editing to localize functional associations in language models~\citep{elhage2021framework, olsson2022induction, meng2022locating, wang2023interp}. The \emph{superposition} hypothesis~\citep{elhage2022superposition} argues that complex features can be distributed across many dimensions, resisting sparse human-readable decomposition. We use representation probing to study information available for pre-generation budget routing in Ref-AVS; within-generation causal mechanisms are outside this evaluation.

\paragraph{Adaptive Compute.}
Adaptive compute has a long history~\citep{zhang2026dypsi}, from adaptive computation time and early-exit networks~\citep{graves2016act, teerapittayanon2016branchynet} to conditional expert routing~\citep{shazeer2017moe}, confidence-based early exiting for BERT (DeeBERT~\citep{xin2020deebert}, FastBERT~\citep{liu2020fastbert}), and autoregressive LMs (CALM~\citep{schuster2022calm}), alongside speculative decoding with small drafter models~\citep{leviathan2023speculative}. Recent LLM/MLLM work adapts reasoning length using certainty estimates, difficulty scores, or controller policies~\citep{wang2025car, xu2025adactrl, tang2025mmecot}. These methods usually evaluate confidence incrementally, depend on external difficulty criteria, or route from surface features. In contrast, our controller makes a single pre-decisional routing decision from \honset, before any reasoning token is generated.

\section{Preliminaries}
\label{sec:prelim}

\paragraph{Ref-AVS Task Formulation.}
Given an untrimmed video $\mathcal{V}$, a synchronized audio track $\mathcal{A}$, and a natural language referring expression $\mathcal{Q}$, Referring Audio-Visual Segmentation predicts a spatio-temporal mask $\mathbf{M} \in \{0, 1\}^{T_v \times H \times W}$ for the sounding target object. Performance is evaluated by region similarity $\mathcal{J}$, contour accuracy $\mathcal{F}$, and their average \jf~\cite{wang2024refavs}. This formulation fixes the downstream objective: all reasoning budgets ultimately differ only in the grounding phrase supplied to the same perception engine, and are judged by the same mask-level metrics. Keeping the perception stack fixed permits comparison of the downstream outcomes induced by the evaluated reasoning policies.

\paragraph{Discrete Reasoning Budgets.}
Following the MLLM pipeline~\citep{tgs2026}, we decouple semantic reasoning from the perception engine, GroundingDINO and SAM2~\citep{liu2023dino, ravi2024sam2}. Keeping the detector and segmenter fixed permits a controlled comparison of the reasoning policies. We define a discrete budget space $\mathcal{B} = \{b_0, b_S, b_L\}$ with cost ordering $b_0 \prec b_S \prec b_L$. The \textbf{Zero} budget $b_0$ bypasses autoregressive reasoning and uses the raw reference $\mathcal{Q}$ as the grounding phrase. The \textbf{Short} budget $b_S$ uses a concise prompt that returns a compact grounding phrase after one reasoning step. The \textbf{Long} budget $b_L$ uses the full TGS-style multimodal CoT prompt before producing the final descriptive phrase. Thus the three budgets define counterfactual policies---no reasoning, limited reasoning, and full reasoning---with the same detector and segmenter applied afterward. Their prompts differ as well as their realized lengths. Token counts are reported in Sec.~\ref{sec:controller_main}, and prompt descriptions are given in Appendix~\ref{app:impl}.

\paragraph{Counterfactual Budget Labeling.}
To define the oracle reasoning need of each sample $s \in \mathcal{S}$, we introduce Counterfactual Budget Labeling (\cbl). Let $\Phi: \mathcal{S} \times \mathcal{B} \to [0, 1]$ map a sample and budget to its empirical \jf score, and let $\Phi^*(s) = \max_{b \in \mathcal{B}} \Phi(s, b)$ be the best achievable performance. We assign $b^*(s)$ as the lowest-cost budget within tolerance $\tau$ of the best score:
$b^*(s) = \min_{\prec} \{ b \in \mathcal{B} \mid \Phi(s, b) \ge \Phi^*(s) - \tau \}$,
where $\tau=0.02$. The tolerance permits a cheaper policy whose score is within 0.02 of the best evaluated outcome; the resulting \cbl policy is cost-aware, not an exact maximum-quality upper bound. These labels supervise probes and the adaptive controller. If all budgets score zero, the rule assigns \textsc{Zero} solely by the cost tie-break. Such \emph{all-fail pipeline cases} provide no successful-budget preference, so the original protocol excludes them from supervision but retains them in every full-raw downstream evaluation. Section~\ref{sec:controller_main} tests their inclusion in supervision directly.

\paragraph{The Generation-Onset Representation.}
Let $\mathbf{X} = [x_1, x_2, \dots, x_N]$ denote the flattened MLLM input sequence containing the prompt, visual context, auditory context, and text query. We define the generation-onset representation as the final-layer state at the last input token, before the first autoregressive reasoning token:
$\honset \triangleq \mathbf{h}_{N}^{(L)}$.
This state is pre-decisional: it has integrated all multimodal evidence but contains no generated reasoning text. It therefore lets us ask whether reasoning need is already encoded before CoT begins, and it can be reused for routing without an additional model forward pass.

\section{Methodology}
\label{sec:method}

\begin{figure}[t]
\centering
\includegraphics[width=\linewidth]{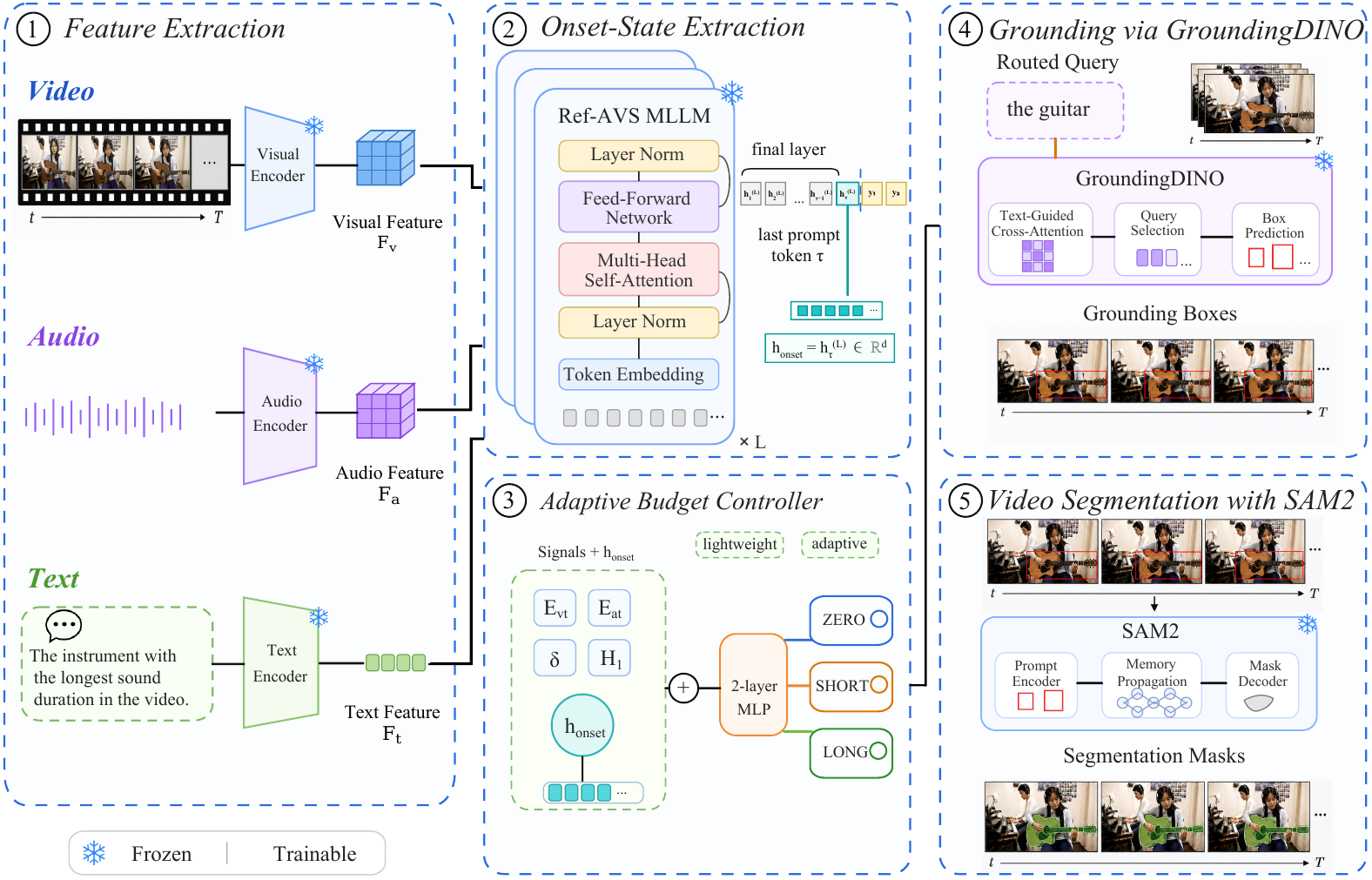}
\caption{\textbf{Pipeline of pre-decisional budget routing for Ref-AVS.} \textbf{(1)} Frozen encoders extract multimodal features. \textbf{(2)} A frozen MLLM integrates these to capture the onset state (\honset) strictly before autoregressive decoding begins. \textbf{(3)} This state drives a lightweight controller to dynamically assign a \textsc{Zero}, \textsc{Short}, or \textsc{Long} reasoning budget. \textbf{(4)} The routed query guides GroundingDINO to predict spatial bounding boxes. \textbf{(5)} These boxes prompt SAM2 to generate the final segmentation masks. Snowflakes denote frozen components; the controller is the sole trainable module.}
\label{fig:pipeline}
\end{figure}

We investigate Ref-AVS reasoning through an analytical and architectural framework centered on the generation-onset representation \honset. As illustrated in Figure~\ref{fig:pipeline}, the same onset state supports representation probing, optional signal analysis, and practical adaptive routing.

\subsection{The Pre-Decisional Budget Signal}
The \pdbs hypothesis is that relative reasoning-budget utility can be predicted from the model state before generation. To test it, we define a readout $f: \mathbb{R}^{d_{model}} \to \mathcal{B}$ that predicts the oracle budget label $b^*(s)$ directly from $\honset(s)$. We deliberately restrict $f$ to a linear classifier, optimizing $\mathbf{W} \in \mathbb{R}^{|\mathcal{B}| \times d_{model}}$ with cross-entropy over $\mathcal{D} = \{(\honset(s), b^*(s))\}_{s \in \mathcal{S}}$. For binary reasoning-need detection, we collapse labels into $b_0$ versus $\{b_S,b_L\}$, i.e., whether reasoning is needed under the evaluated \cbl rule. Successful linear readout would establish predictive information available before active reasoning begins.

\subsection{Hand-Crafted Signals: A Proxy Comparison}
To test whether \pdbs is governed by simple heuristics, we construct four interpretable, modality-aware signals. If budget selection were reducible to explicit uncertainty or alignment cues, these low-dimensional factors should approximate the predictive power of \honset.

\paragraph{Cross-modal attention entropy.} High attention entropy from textual tokens to visual or auditory tokens suggests diffuse and uncertain modality grounding. Let $\mathcal{R}$ be the set of reference-token positions and let $\bar{a}^{(\ell,h)}_{i\to V}$ denote attention weights restricted to visual tokens and renormalized within that subset. We aggregate visual attention entropy over all layers, heads, and reference tokens:
\begin{equation}
E_{vt}(s) = (L H |\mathcal{R}|)^{-1} \sum_{\ell=1}^{L} \sum_{h=1}^{H} \sum_{i\in\mathcal{R}} \mathcal{H}\big( \bar{a}^{(\ell,h)}_{i\to V} \big),
\end{equation}
where $\mathcal{H}(\cdot)$ denotes the Shannon entropy function. The auditory counterpart $E_{at}$ is defined analogously over renormalized audio-token attention.

\paragraph{Latent contextual shift.} Let \hpre be the final-layer state at the last audio/video context token before reference-text tokens are consumed. We measure the magnitude of the contextual change induced by the natural language query as $\delta(s) = \|\honset(s) - \hpre(s)\|_{2}$.

\paragraph{First-token decoding uncertainty.} To capture the residual ambiguity at the absolute onset of generation, we compute the entropy over renormalized top-five next-token probabilities:
$H_1(s) = - \sum_{v \in \text{top-5}} \tilde{p}(v) \log \tilde{p}(v)$, where $\tilde{p}(v)=p(v)/\sum_{u\in\text{top-5}}p(u)$.

\paragraph{The proxy comparison.} We compare readouts from the joint feature vector $\mathbf{c} = [E_{vt}, E_{at}, \delta, H_1]^\top \in \mathbb{R}^4$ with readouts from the full \honset vector. A performance gap measures predictive information missed by these evaluated summaries; it does not rule out all possible compact feature families.

\subsection{Projection Removal Intervention}
To test whether the readable budget direction is useful for routing, we perform a targeted projection-removal ablation. We estimate the primary budget-predictive direction $\mathbf{w}_{budget}$ from the binary probe on training folds only, apply the projection to both training and held-out onset features, and retrain a fresh probe on the projected training features. This tests readout degradation after refitting rather than only misalignment with fixed classifier weights. It does not modify the frozen MLLM's generation activations.

\subsection{The Single Forward-Pass Controller}
Building on the predictive onset-state readout, we design a single forward-pass controller for dynamic reasoning-budget routing. The primary controller is an ultra-lightweight two-layer multilayer perceptron (MLP) that maps $\honset \in \mathbb{R}^{d_{model}}$ directly to the discrete budget space $\mathcal{B}$. Figure~\ref{fig:pipeline} also shows the hand-crafted signal branch, $\mathbf{c} = [E_{vt}, E_{at}, \delta, H_1]^\top$, because it is used for proxy analysis and ablation; all main controller results use \honset alone.

The routing probability is computed with a softmax over the MLP output, and the query is dispatched by $\hat{b} = \arg\max_b P(b|s)$. Unlike cascaded systems or external gating networks, the controller reuses the same forward pass the backbone already performs to encode the prompt. The decision is finalized at the generation onset, preserving architectural minimalism and adding near-zero latency.
\section{Experiments}
\label{sec:exp}

\subsection{Experimental Setup}
\label{sec:exp_setup}
\textbf{Backbone \& Data.} We instantiate our multimodal reasoning framework using a Llama-2-7B backbone augmented by LoRA adapters~\cite{hu2021lora}, adopting the architecture from~\cite{tgs2026}. We use merged LoRA weights for inference efficiency; the reported implementation check has approximately 3\% relative hidden-state norm difference. Main evaluations use RefAVSBench~\cite{wang2024refavs} test\_s (2,287 samples) and test\_u (1,646 samples), with source-only transfer to R$^2$-AVSBench~\cite{tgs2026} test\_s (219 samples). The original splits use seed 42, with confidence intervals from 1,000 bootstrap resamples and probing significance from exact binomial tests. Further details are in Appendix~\ref{app:impl}.

RefAVSBench readout and router metrics are five-fold out-of-fold estimates within these benchmark populations: each sample is predicted by a readout trained without that sample, with validation confined to the training folds. They are not auxiliary-controller evaluations on an untouched official test split. All downstream \jf and token metrics retain the complete raw populations. The original R$^2$ result uses no target-label fitting; because its references are rewritten over RefAVS videos, it tests reference shift rather than source-video-disjoint transfer. Group-disjoint native adaptation is evaluated separately in Appendix~\ref{app:calibration}.

\subsection{Quantifying Reasoning-Budget Heterogeneity}
\label{sec:exp_het}

\begin{table}[h]
\caption{\textbf{Budget heterogeneity across benchmarks.} Budget columns are ordered \textsc{Zero}/\textsc{Short}/\textsc{Long}. Shares use $N_{\text{CBL}}$, excluding all-fail pipeline cases from original supervision; \jf uses all $N_{\text{raw}}$ cases. The cost-aware \cbl oracle exceeds always-long by $+5.3$ to $+7.4$ \jf points.}
\label{tab:heterogeneity}
\centering\small
\setlength{\tabcolsep}{4pt}
\begin{tabular}{@{}lrrccc ccc@{}}
\toprule
\multirow{2}{*}{Split} & \multirow{2}{*}{$N_{\text{raw}}$} & \multirow{2}{*}{$N_{\text{CBL}}$} & \multicolumn{3}{c}{Budget distribution (\%)} & \multicolumn{2}{c}{\jf} & \multirow{2}{*}{$\Delta$} \\
\cmidrule(lr){4-6} \cmidrule(lr){7-8}
 & & & zero & short & long & oracle & a-long & \\
\midrule
RefAVSBench test\_s    & 2287 & 2122 & 42.4 & 27.3 & 30.3 & 0.623 & 0.549 & $+$0.074 \\
RefAVSBench test\_u    & 1646 & 1578 & 57.7 & 24.1 & 18.2 & 0.822 & 0.769 & $+$0.053 \\
R$^2$-AVSBench test\_s &  219 &  204 & 41.1 & 32.4 & 26.5 & 0.539 & 0.474 & $+$0.065 \\
\bottomrule
\end{tabular}
\end{table}

Table~\ref{tab:heterogeneity} motivates the need for adaptive reasoning budgets. No fixed budget is preferred for every sample; even the largest single-budget label share remains below 60\%. The gap between oracle assignment and always-long reasoning reaches 5.3--7.4 \jf points---a large margin in Ref-AVS, where adaptive architectures improve over prior methods by roughly 4 points~\cite{tgs2026}. All three budgets receive substantial label mass on every split, indicating recurring budget heterogeneity rather than a single-split artifact.

Specifically, a one-size-fits-all \emph{always-long} strategy can waste computation and harm segmentation. The 42.4\% \emph{need-zero} share in RefAVSBench test\_s identifies cases where direct grounding is within $\tau$ of the best evaluated outcome; it does not mean that long reasoning harms every such case, or that their absolute quality is uniformly high. The harmful-long subset is examined below. Conversely, the 30.3\% \emph{need-long} share contains cases where neither cheaper policy reaches the tolerance band. The presence of all three labels on R$^2$-AVSBench further motivates testing a pre-generation signal for budget selection beyond the in-distribution reference set.

\paragraph{Characterizing the overthinking trap.}
To characterize the overthinking trap, we randomly sample 200 \emph{need-zero} examples satisfying $\Phi(s,b_0)-\Phi(s,b_L)>\tau$. Two annotators, blind to the budget identity, assess whether the generated phrase introduces unsupported attributes, incorrect object categories, or hallucinated spatial relations not present in the video. The reported inter-annotator agreement is Cohen's $\kappa=0.82$; disagreements are resolved in a third pass. This outcome-selected audit characterizes harmful-long cases, not the prevalence of these errors in the complete need-zero class.

On these outcome-selected examples, \textsc{Long} CoT increases grounding phrase length by 245\%, and 41.5\% of its outputs contain unsupported additions (Appendix Table~\ref{tab:error_analysis}). This metric is not applicable to \textsc{Zero}, which uses the original reference rather than generating a refined phrase. Alongside these phrase-level changes, top-1 box intersection over union (IoU) drops from 0.72 to 0.58, and \jf falls by 16.5 points. These paired observations illustrate harmful reasoning on the audited subset without isolating a causal mediation pathway.
\subsection{Pre-Decisional Readout of Reasoning Need}
\label{sec:exp_probe}

\begin{figure}[t!]
\centering
\includegraphics[width=0.9\linewidth]{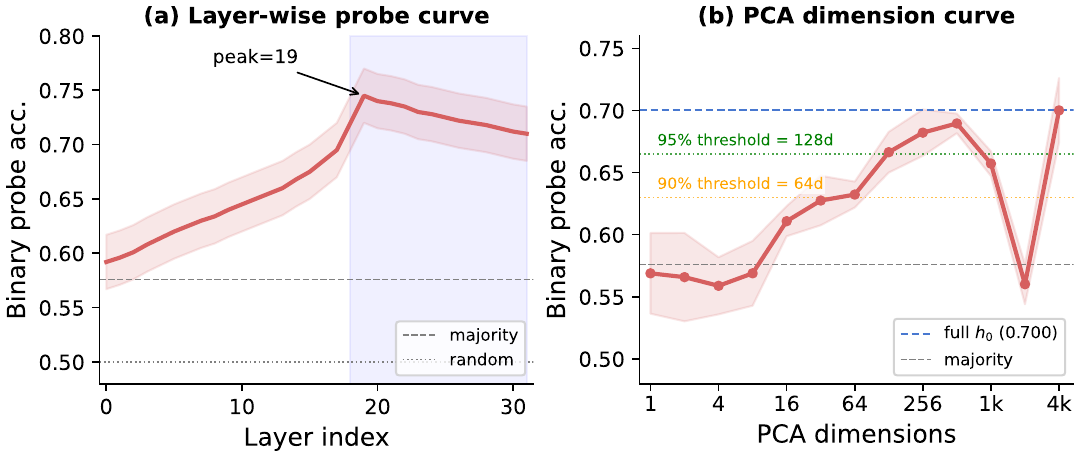}
\caption{\textbf{\pdbs is readable across layers and dimensions.} Layer-wise readability peaks at L19 and remains informative at the final layer. The principal component analysis (PCA) curve characterizes readout accuracy under the evaluated projection procedure; recovery in a reduced space does not establish an irreducibly high-dimensional encoding.}
\label{fig:encodability_main}
\end{figure}

\begin{table}[htbp]
\caption{\textbf{Linear probes on \honset exceed baselines.} Baselines are matched to the same 5-fold split. Binary accuracy collapses labels into $b_0$ versus $\{b_S,b_L\}$, and significance ($p_{\text{bin}}$) is tested against the binary majority baseline. Linear readout is strongest for binary reasoning-need detection, while the MLP improves 3-way budget classification. PBS-A denotes MAR3's text-only reference router.}
\label{tab:probes}
\centering\small
\setlength{\tabcolsep}{4pt}
\begin{tabular}{@{}lccc@{}}
\toprule
Probe input & Bin.\ acc $\uparrow$ & 3-cls acc $\uparrow$ & $p_{\text{bin}}$ vs.\ majority \\
\midrule
Majority class                          & 0.577 & 0.424 & ---  \\
Random Gaussian ($d{=}4096$)            & 0.577\,{\scriptsize$\pm$.024} & 0.425\,{\scriptsize$\pm$.018} & $>0.4$ \\
Text only (PBS-A)~\citep{mar3_2026}     & 0.581 & 0.431 & 0.68  \\
Ref-token embed.\ (mean)                & 0.623 & 0.460 & $<10^{-5}$  \\
\midrule
\textbf{Linear on \honset}              & \textbf{0.701}\,{\scriptsize$\pm$.026} & 0.503 & $<10^{-30}$ \\
MLP on \honset (2-layer)                & 0.689 & \textbf{0.546}\,{\scriptsize$\pm$.026} & $<10^{-25}$ \\
Linear, best layer (L19)                & \textbf{0.745} & 0.538 & --- \\
\bottomrule
\end{tabular}
\end{table}

As shown in Table~\ref{tab:probes}, \pdbs is linearly readable before reasoning starts. A single linear layer on \honset achieves 70.1\% binary accuracy, exceeding the 57.7\% majority baseline ($p < 10^{-30}$). Because \honset is captured at generation onset, this signal is not derived from generated reasoning text; it is measured before autoregressive reasoning begins. 

By the final prompt token, the model has integrated multimodal evidence into a state that contains information predictive of reasoning need, whereas surface properties such as text-only features (58.1\%) fall substantially behind. Figure~\ref{fig:encodability_main} shows that this readability is not localized to a single layer: it peaks at L19 and remains informative at the final layer. The dimensional recovery curve shows that substantial readout performance is retained in a reduced PCA space, without identifying a unique encoding mechanism.

\paragraph{Balanced readout and label separation.}
The binary linear probe achieves 70.0\% balanced accuracy, with 69.2\% need-zero recall and 70.8\% need-reasoning recall, versus 50.0\% balanced accuracy for a constant majority predictor. The deployed three-class router, collapsed to binary, has 66.5\% accuracy and 66.3\% balanced accuracy; it differs from the separately trained binary MLP at 68.9\%. The dimension-matched Gaussian control remains near the majority baseline, showing that dimensionality alone does not explain the gain under this protocol.

To examine short-class errors, we define $m(s)=\Phi^*(s)-\max_{b\ne b^*(s)}\Phi(s,b)$. A within-tolerance \cbl choice that is not a strict maximizer has $m(s)=0$. A median split of the 2,122 informative cases gives high-separation/near-tie three-class accuracy 0.641/0.451 and need-short recall 0.542/0.178. These results associate short-class errors with near-tied outcomes; they do not establish an intrinsically ambiguous transition class. All trajectories use fixed greedy decoding; the stratification does not establish stochastic label noise.

\paragraph{Cross-backbone readability.}
To test whether onset-state readability is unique to our primary Llama-2-7B checkpoint, we run a lightweight cross-backbone probe study. We extract generation-onset states from identically prompted alternative architectures, including Vicuna-7B and Qwen-VL. Qwen-VL is a vision-language architecture, so we augment its prompt with audio-derived features for modality matching. Rather than generating new counterfactual labels for this probe comparison, we transfer the \cbl labels from the primary TGS-Agent pipeline. Table~\ref{tab:cross_backbone} measures \emph{representation concordance} with those primary-pipeline labels, not accuracy against each backbone's native optimal budgets. Label mismatch can increase or decrease this concordance. Separate native-label calibration experiments appear in Appendix~\ref{app:calibration}.

\begin{table}[htbp]
\centering\small
\caption{\textbf{Cross-backbone probe readability.} Alternative MLLM onset states can predict the reasoning-utility structure defined by our primary counterfactual labeling protocol. The majority baseline is identical because all rows are evaluated against the same binary CBL label distribution.}
\label{tab:cross_backbone}
\setlength{\tabcolsep}{4pt}
\begin{tabular}{lllccc}
\toprule
\textbf{Backbone} & \textbf{Modalities} & \textbf{Label source} & \textbf{Bin.\ acc.} & \textbf{Majority} & \textbf{$p$-value} \\
\midrule
Llama-2-7B (primary) & A / V / T & Own CBL & 0.701 & 0.577 & $<10^{-30}$ \\
Vicuna-7B & A / V / T & Transferred CBL & 0.692 & 0.577 & $<10^{-20}$ \\
Qwen-VL & V / T + Audio feats. & Transferred CBL & 0.704 & 0.577 & $<10^{-20}$ \\
\bottomrule
\end{tabular}
\end{table}

\subsection{\pdbs Compared with Interpretable Proxies}
\label{sec:exp_decomp}

The evaluated hand-crafted signals do not fully recover the predictive information in \honset. Appendix Figure~\ref{fig:distributed} reports near-chance areas under the receiver operating characteristic curve (AUCs) for individual signals and compares the original four-signal readouts. Under the same five-fold, class-balanced MLP protocol, four signals give 0.541 binary accuracy. Adding first-token logit margin gives 0.552; adding a per-layer entropy profile gives 0.578; combining both gives 0.586. Stronger features thus narrow the gap to the onset MLP's 0.689 from 14.8 to 10.3 points. This supports a claim limited to the \emph{evaluated proxies}, not irreducibility, superposition, or a causal mechanism. Detailed values are in Appendix~\ref{app:signal_table}.

\subsection{Functional Evidence for the Routing Subspace via Projection Removal}
\label{sec:exp_patch}

While linear readability establishes correlation, we test whether the decoded \pdbs direction is useful for the external router. We identify the budget-predictive direction $\mathbf{w}_{budget} \in \mathbb{R}^{d_{model}}$ from a binary probe fitted only on training folds. The same training-derived projection is applied to training and held-out features. For a given onset state $\mathbf{h}$, we ablate this direction:
$\mathbf{h}_{removed} = \mathbf{h} - (\mathbf{w}_{budget}^{\top}\mathbf{h}/\|\mathbf{w}_{budget}\|^2)\mathbf{w}_{budget}$.
We then retrain a fresh linear probe on the projected training features and evaluate on the corresponding held-out fold.

As shown in Appendix Table~\ref{tab:projection}, retrained-probe accuracy drops from 70.1\% to 61.2\%, while routed \jf drops from 0.528 to 0.480. Under 1,000 paired bootstrap resamples, these differences are $-0.089$, with a 95\% confidence interval (CI) of $[-0.103,-0.075]$, and $-0.048$ $[-0.056,-0.040]$, respectively. Random-direction removal changes routed \jf by $-0.003$ $[-0.008,0.002]$. The selected direction is therefore useful to the external readout/routing pipeline. Because this ablation modifies readout features rather than generation activations, it does not establish within-model causal self-assessment.
\subsection{Single Forward-Pass Controller Performance}
\label{sec:exp_ctrl}
\label{sec:controller_main}

To evaluate our \pdbs controller, we compare against static policies and four adaptive baselines: a text-only PBS-A router, a multimodal mean-pooling pre-token router, a \textbf{Confidence Router} based on GroundingDINO top-1 box scores after an initial \textsc{Zero} perception pass, and a \textbf{Short-then-Decide} pipeline that generates \textsc{Short} reasoning before deciding whether to upgrade to \textsc{Long}.

\begin{table}[htbp]
\centering\scriptsize
\caption{\textbf{Comprehensive comparison of routing strategies.} Bold marks the highest \jf among adaptive methods and the lower token cost of onset routing. All raw cases are included; intervals use 1,000 paired bootstraps. Out-of-distribution (OOD) results use source-only transfer to rewritten references, without target-label fitting. Token columns report in-distribution (ID) costs only.}
\label{tab:main_routing}
\setlength{\tabcolsep}{3pt}
\resizebox{\textwidth}{!}{%
\begin{tabular}{llcccc}
\toprule
\textbf{Method} & \textbf{Decision point} & \textbf{ID \jf} & \textbf{OOD \jf} & \textbf{Avg tok. (ID)} & \textbf{Relative tok. (ID)} \\
\midrule
\multicolumn{6}{l}{\textit{Static budgets and cost-aware oracle}} \\
Always-Zero & Fixed (no CoT) & 0.338 $\pm$ 0.012 & 0.254 $\pm$ 0.038 & 0 & 0.0\% \\
Always-Short & Fixed & 0.434 $\pm$ 0.015 & 0.394 $\pm$ 0.035 & 93 & 33.8\% \\
Always-Long~\cite{tgs2026} & Fixed (full CoT) & 0.549 $\pm$ 0.016 & 0.474 $\pm$ 0.033 & 275 & 100.0\% \\
Oracle \cbl & Oracle & 0.623 $\pm$ 0.014 & 0.539 $\pm$ 0.031 & 109 & 39.6\% \\
\midrule
\multicolumn{6}{l}{\textit{Adaptive baselines}} \\
Text Router (PBS-A)~\cite{mar3_2026} & Pre-token & 0.411 $\pm$ 0.018 & 0.358 $\pm$ 0.034 & 57 & 20.7\% \\
Multimodal Mean Pool & Pre-token & 0.445 $\pm$ 0.017 & 0.380 $\pm$ 0.036 & 105 & 38.1\% \\
Confidence Router & Post-\textsc{Zero} detector pass & 0.467 $\pm$ 0.016 & 0.401 $\pm$ 0.035 & 145 & 52.7\% \\
Short-then-Decide & Post-\textsc{Short} CoT & \textbf{0.535} $\pm$ \textbf{0.015} & \textbf{0.461} $\pm$ \textbf{0.032} & 182 & 66.1\% \\
\midrule
\multicolumn{6}{l}{\textit{Ours}} \\
\textbf{\honset Router (Ours)} & \textbf{Pre-token (\pdbs)} & 0.528 $\pm$ 0.015 & 0.458 $\pm$ 0.032 & \textbf{110} & \textbf{40.0\%} \\
\bottomrule
\end{tabular}%
}
\end{table}

The onset controller lies on the evaluated Pareto frontier (Table~\ref{tab:main_routing}). Relative to \emph{Short-then-Decide}, reasoning tokens are reduced by 39.6\% with a 0.007 decrease in \jf; relative to the \emph{Confidence Router}, \jf increases by 0.061 without a detector pass before routing. Source-only transfer to R$^2$-AVSBench yields 0.458 \jf, below Always-Long's 0.474, so the OOD benefit is cost-efficiency rather than accuracy. The budget is selected from fused multimodal context before either reasoning generation or downstream perception is required.

\paragraph{Full-raw robustness.}
All 165 all-fail inputs remain in every full-raw test\_s evaluation. Adding their formal cost-tie \textsc{Zero} labels to supervision changes \jf from 0.5280 to 0.5278 (difference $-0.0002$, 95\% CI $[-0.0018,0.0014]$) and tokens from 110.0 to 105.3 ($-4.7$ $[-5.6,-3.8]$), under the same folds and configuration. Thus, excluding them from original supervision does not drive the reported quality--cost point under this protocol. On these cases, mean tokens fall from 116 to 51 while every evaluated action still scores zero; their route affects cost but not \jf directly.

\paragraph{Measured decision and inference latency.}
A paired batch-one A100 benchmark over all 2,287 inputs gives mean decision times of 0.82 versus 2.13\,s and MLLM latencies of 2.31 versus 3.17\,s for onset routing versus \emph{Short-then-Decide}. The paired differences are $-1.31$\,s $[-1.34,-1.28]$ and $-0.86$\,s $[-0.94,-0.78]$. With prefix-cache reuse, \emph{Short-then-Decide} takes 2.92\,s, still 0.61\,s slower $[0.53,0.69]$. These measurements quantify the advantage when early decisions or lower mean inference latency matter; they do not imply quality dominance over the original 0.535 baseline. Full timing and sensitivity protocols are provided in the appendix.

\paragraph{Calibration cost and finer action spaces.}
Further experiments test two deployment constraints without replacing the original operating point. Using 25\% of source calibration labels gives 0.527 \jf at 110 tokens, versus 0.530 at 109 for matched full-data retraining. Under group-disjoint R$^2$ adaptation, 20\% native calibration reaches 0.470 versus 0.471 for full calibration; native-only Qwen calibration reaches 0.536 versus 0.538. The fold-matched upper break-even estimates fall from 8,910 to 2,310 source-domain queries and from 10,180 to 2,070 native-Qwen queries. These savings require the stated accounting: retaining transferred labels is not a cheaper cold start when constructing their table is charged (Appendix~\ref{app:calibration}).

With nested semantic action sets, matched Utility-3/5/7 controllers reach 0.531/0.536/0.540 \jf at 110/111/112 tokens. Utility-7 exceeds Utility-3 by 0.009, with Holm-adjusted 95\% CI $[0.0054,0.0126]$. Its 0.540/112 operating point also improves on Short-then-Decide's 0.535/182 point estimates; this latter comparison is not a separate paired significance test. A same-checkpoint action-mask comparison confirms a 0.006 gain $[0.0031,0.0089]$ from making the added actions available without refitting. These results demonstrate extensibility within the evaluated policy family, not optimality over arbitrary token- or layer-level exits (Appendix~\ref{app:actions}).

Appendix Figure~\ref{fig:qualitative_main} illustrates the same routing behavior qualitatively; Appendix Figures~\ref{fig:pareto}--\ref{fig:perlabel} and Appendix~\ref{app:sota} provide cost, per-label, and SOTA diagnostics.
\section{Discussion \& Limitations}
\label{sec:discuss}

Linear readability, proxy comparisons, and projection-removal effects support the utility of \honset for external Ref-AVS routing. These results concern readout features rather than the frozen MLLM's generation activations. They therefore motivate efficient pre-generation routing without establishing causal self-assessment by the backbone.

\cbl requires target masks and pipeline-specific outcomes, so material changes to the backbone, prompts, or input distribution may require recalibration. Offline costs must be amortized over repeated queries. The three-policy action set provides an auditable, rather than optimal, discretization. Appendices~\ref{app:calibration}--\ref{app:actions} evaluate reduced native calibration and nested five-/seven-action policies; these results do not establish prompt-invariant calibration or arbitrary token-/layer-level early exiting. Wider task transfer and within-generation causal mechanisms remain open questions.

\section{Conclusion}
Pre-generation budget routing is studied in Ref-AVS through \pdbs, the predictive information encoded in generation-onset hidden states. A lightweight controller reads this signal to bypass unnecessary reasoning and allocate longer reasoning when useful. Linear probes, stronger proxy comparisons, projection-removal ablations, and full-raw evaluations support its utility for external routing within the evaluated pipelines. The original three-action controller retains approximately 96\% of Always-Long \jf while reducing reasoning tokens by 60\%. Source-only transfer provides a further cost--quality trade-off. Together, these findings identify generation onset as a useful decision point for allocating reasoning budgets.
\begin{ack}
This work was supported in part by the New Generation Artificial Intelligence-National Science and Technology Major Project under Grant 2025ZD0123701, in part by the National Natural Science Foundation of China under Grants 62672348, 62476202, and 62272343, in part by the Explorers Program of Shanghai (Basic Research Funding) from Shanghai Municipal Science and Technology Commission under Grant 26TS1403800, in part by the Innovation Ecosystem Development Program of Shanghai Municipal Science and Technology Commission under Grant 26510761200, and in part by the Fundamental Research Funds for the Central Universities.
The authors declare no competing interests.
\end{ack}

\clearpage
\bibliographystyle{plainnat}

\appendix

\section{Full RefAVSBench Comparison}
\label{app:sota}

Table~\ref{tab:comparison} situates our adaptive controller within the broader landscape of Ref-AVS benchmarks. Baseline entries are reproduced from the TGS-Agent comparison table~\cite{tgs2026}; they are not all independently rerun under the auxiliary controller's out-of-fold protocol. We append our budget-routing policies. At 40\% of the Always-Long reasoning-token budget, the original adaptive controller achieves 52.8 \jf on the Seen split.

The reported component scores reveal a trade-off: the adaptive controller has higher $\mathcal{J}$ on Seen and Mix and slightly higher $\mathcal{F}$ on Unseen than the reported TGS-Agent baseline, while its overall \jf is lower. These aggregate differences do not identify the mechanism responsible. The Mix summaries follow the benchmark comparison and should not be interpreted as sample-count-weighted averages of the displayed Seen and Unseen scores.

\begin{table}[h]
\caption{\textbf{Quantitative comparison on the RefAVSBench dataset.} We report the region Jaccard index ($\mathcal{J}$), contour accuracy ($\mathcal{F}$), and their average (\jf). ``Null'' denotes the absent-target false-positive score used by RefAVSBench, where lower is better; ``Tok (S)'' denotes average autoregressive reasoning tokens on the Seen split. Baseline rows, including the identical GAVS/SAMA values, follow the TGS-Agent comparison table. Best practical results, excluding Oracle, are highlighted in \textbf{bold}.}
\label{tab:comparison}
\centering\scriptsize
\setlength{\tabcolsep}{3pt}
\resizebox{\linewidth}{!}{%
\begin{tabular}{@{}lllc ccc ccc ccc cc@{}}
\toprule
\multirow{2}{*}{Method} & \multirow{2}{*}{Venue} & \multirow{2}{*}{Task} &
\multicolumn{3}{c}{Seen (S) $\uparrow$} &
\multicolumn{3}{c}{Unseen (U) $\uparrow$} &
\multicolumn{3}{c}{Mix (S+U) $\uparrow$} &
\multirow{2}{*}{Null $\downarrow$} & \multirow{2}{*}{Tok (S)} \\
\cmidrule(lr){4-6}\cmidrule(lr){7-9}\cmidrule(lr){10-12}
 & & & $\mathcal{J}$ & $\mathcal{F}$ & \jf & $\mathcal{J}$ & $\mathcal{F}$ & \jf & $\mathcal{J}$ & $\mathcal{F}$ & \jf & & \\
\midrule
AVSBench~\cite{zhou2022avs}           & ECCV'22 & AVS     & 23.2 & 51.1 & 37.2 & 32.4 & 54.7 & 43.5 & 27.8 & 52.9 & 40.3 & 0.208 & -- \\
AVSegFormer~\cite{gao2024avsegformer} & AAAI'24 & AVS     & 34.5 & 47.0 & 40.8 & 36.1 & 50.1 & 43.1 & 34.8 & 48.6 & 41.7 & 0.171 & -- \\
GAVS~\cite{wang2024gavs}              & AAAI'24 & AVS     & 28.9 & 49.8 & 39.4 & 29.8 & 49.7 & 39.8 & 29.4 & 49.8 & 39.6 & 0.190 & -- \\
SAMA~\cite{liu2024sama}               & WACV'24 & AVS     & 28.9 & 49.8 & 39.4 & 29.8 & 49.7 & 39.8 & 29.4 & 49.8 & 39.6 & 0.190 & -- \\
\midrule
ReferFormer~\cite{wu2022referformer}  & CVPR'22 & Ref-VOS & 31.3 & 50.1 & 40.7 & 30.4 & 48.8 & 39.6 & 30.9 & 49.5 & 40.2 & 0.176 & -- \\
R2VOS~\cite{li2023r2vos}              & ICCV'23 & Ref-VOS & 25.0 & 41.0 & 33.0 & 27.9 & 49.8 & 38.9 & 26.5 & 45.4 & 35.9 & 0.183 & -- \\
\midrule
EEMC~\cite{wang2024refavs}            & ECCV'24 & Ref-AVS & 34.2 & 51.3 & 42.8 & 49.5 & 64.8 & 57.2 & 41.9 & 58.1 & 50.0 & \textbf{0.007} & -- \\
Grounded-SAM2~\cite{ren2024gdsam2}    & arXiv'24 & Ref-AVS & 28.5 & 39.9 & 34.2 & 59.8 & 68.1 & 63.9 & 44.2 & 54.0 & 49.1 & 0.277 & -- \\
Crab~\cite{du2025crab}                & CVPR'25 & Ref-AVS & 40.5 & -- & -- & 45.6 & -- & -- & 43.1 & -- & -- & -- & -- \\
TSAM~\cite{radman2025tsam}            & CVPR'25 & Ref-AVS & 43.4 & 56.8 & 50.1 & 54.6 & 66.4 & 60.5 & 49.0 & 61.6 & 55.3 & 0.017 & -- \\
SAM2-LOVE~\cite{wang2025sam2love}     & CVPR'25 & Ref-AVS & 43.5 & 51.9 & 47.7 & 66.5 & 72.3 & 69.4 & 55.0 & 62.1 & 58.5 & 0.230 & -- \\
TGS-Agent~\cite{tgs2026}              & AAAI'26 & Ref-AVS & 49.5 & \textbf{60.4} & \textbf{54.9} & \textbf{73.2} & 80.6 & \textbf{76.9} & 61.3 & \textbf{70.5} & \textbf{65.9} & 0.035 & 275 \\
\midrule
\textbf{Ours} (adaptive \honset)        & --      & Ref-AVS & \textbf{56.0} & 49.6 & 52.8 & 66.8 & \textbf{80.8} & 73.8 & \textbf{61.4} & 65.2 & 63.3 & 0.033 & \textbf{110} \\
\cbl policy oracle                     & --      & Ref-AVS & 58.0 & 66.6 & 62.3 & 78.4 & 86.0 & 82.2 & 68.2 & 76.3 & 72.3 & 0.015 & 109 \\
\bottomrule
\end{tabular}
}
\end{table}

\section{Implementation Details}
\label{app:impl}

\paragraph{LoRA merge verification.}
We use merged LoRA weights throughout the primary probing and routing experiments. An implementation check on sample \texttt{-3ABOVeVmpU\_136000\_146000\_piano\_1} gives relative feature difference $\|\honset^{\text{LoRA-direct}}-\honset^{\text{merged}}\|/\|\honset\|=0.030$. This single-sample check documents the observed discrepancy rather than establishing population-wide equivalence of the representations.

\paragraph{Position locator.}
Accurate extraction of \honset relies on precise token indexing. Let $t_{\text{pre}}$ and $t_0$ denote the final audio/video context token index and final reference-text token index, respectively. These extraction boundaries were manually validated across samples with highly variable reference lengths to check that the generation-onset state is captured before decoding begins.

\paragraph{Probe and controller training.}
Both linear probes and MLP controllers are optimized using Adam with a learning rate of $lr=10^{-3}$ and a batch size of 32. We train for a maximum of 50 epochs, employing early stopping based on validation loss. The MLP architecture maps $4096 \to 256 \to 3$ with ReLU activations and a 0.1 dropout rate. For controller training, we utilize class-balanced cross-entropy weights to account for dataset imbalances (e.g., the \emph{need-short} class represents about 25\% of the distribution). The optional ablation controller concatenates the hand-crafted signals, yielding the 4,100-dimensional input $[\honset;\Evt;\Eat;\delt;\Hone]$. RefAVSBench estimates use a 5-fold cross-validation scheme seeded at 42; the original source-only R$^2$-AVSBench evaluation uses no target-label training.

\paragraph{Reasoning-budget prompts.}
The \textsc{Zero} budget skips autoregressive decoding entirely and directly sends $\mathcal{Q}$ to the grounding module. \textsc{Short} and \textsc{Long} share the same multimodal inputs and stop-token template as TGS-Agent, but differ in instruction: \textsc{Short} asks for a concise grounding phrase after one reasoning step, whereas \textsc{Long} uses the full object-aware Think-Ground-Segment prompt. We report realized average token counts rather than treating prompt budgets as hard token caps. Original \cbl trajectories use greedy decoding (\texttt{do\_sample=False}; sampling temperature is inactive). The onset router still performs multimodal input encoding when it selects \textsc{Zero}.

\paragraph{Out-of-fold evaluation protocol.}
All prompts, the \cbl tolerance $\tau=0.02$, controller architectures, and hyperparameters are fixed for the original evaluation. For each five-fold split, router training uses only CBL labels from the training folds. Early stopping uses a validation subset drawn exclusively from those folds; held-out labels are used only for metric reporting. The RefAVSBench results are therefore out-of-fold estimates within the benchmark populations, not evaluations of an auxiliary controller trained exclusively on an official training split and tested on an untouched test population. For projection removal, $\mathbf{w}_{budget}$ is estimated on training folds; a fresh probe is trained on projected training states and evaluated on projected held-out states. The original R$^2$-AVSBench evaluation uses no target-label fitting or tuning. Since R$^2$ references are rewritten over RefAVS videos, it does not by itself establish source-video-disjoint transfer. Native target calibration in Appendix~\ref{app:calibration} is a separate experiment.

\paragraph{Adaptive baseline implementation.}
To ensure a fair comparison, all adaptive baselines are trained under the identical 5-fold cross-validation protocol. The \emph{Multimodal Mean Pool} baseline averages the final-layer visual, audio, and reference-text token states to train a matching two-layer MLP controller. The \emph{Confidence Router} extracts the GroundingDINO top-1 box score from an initial \textsc{Zero} perception pass and routes it through a class-balanced MLP. The \emph{Short-then-Decide} baseline forces every sample through a \textsc{Short} generation sequence, subsequently utilizing the aggregated hidden and logit features of the \textsc{Short} output to train a secondary router that decides whether to upgrade to a \textsc{Long} budget.

\paragraph{Cross-backbone audio interface.}
For architectures such as Qwen-VL that do not natively tokenize raw audio streams, we supply audio-derived information through an audio-summary prompt and feature interface. The original cross-backbone experiments are probe-only evaluations using transferred CBL labels from the primary TGS-Agent pipeline. Their accuracies measure concordance with that label source, not native optimal-budget accuracy; label mismatch may increase or decrease concordance. The later native-label Qwen experiments in Appendix~\ref{app:calibration} address a separate downstream calibration question.

\paragraph{Hardware and computational footprint.}
The primary feature extraction, probe training, and inference routing experiments used a single NVIDIA A100 (80GB) GPU. The original study recorded approximately 85 GPU-hours and 10 CPU-hours for data processing and analysis. These are original-study totals, not a complete aggregate of all later extension experiments; measured setup costs for those experiments are reported separately in Appendix~\ref{app:calibration}.

\section{CBL Threshold Sweep}
\label{app:cbl_sweep}

The sensitivity of \cbl to its quality tolerance is evaluated over $\tau \in [0, 0.03]$. The three-class budget-label distribution remains stable across these settings. Cohen's $\kappa$ is at least 0.79 relative to the default $\tau=0.02$, indicating that the observed heterogeneity persists across the evaluated tolerances rather than depending on one threshold.

\vspace{3pt}
\begin{center}\small
\begin{tabular}{cccc|c}
\toprule
$\tau$ & need-zero & need-short & need-long & Cohen's $\kappa$ vs.\ $\tau{=}0.02$ \\
\midrule
0.00  & 38.2\% & 29.6\% & 32.2\% & 0.79 \\
0.01  & 40.3\% & 28.5\% & 31.2\% & 0.88 \\
\textbf{0.02} & \textbf{42.4\%} & \textbf{27.3\%} & \textbf{30.3\%} & \textbf{1.00} (ref) \\
0.03  & 44.7\% & 25.9\% & 29.4\% & 0.91 \\
\bottomrule
\end{tabular}
\end{center}

\section{Additional Robustness Checks}
\label{app:extra}

\paragraph{Attention aggregation strategies.}
Computing cross-modal attention entropy (\Evt, \Eat) over only the last four layers, or exclusively over the five most variable attention heads, yielded binary AUC scores below 0.53. These checks show limited discrimination for the tested aggregations; they do not rule out other attention summaries.

\paragraph{Audio modality contribution.}
Replacing the raw audio input with Whisper-generated ASR transcripts reduces \jf by 0.006 on test\_s and 0.008 on test\_u. The evaluated audio input therefore provides information useful beyond this transcript interface, although the comparison does not identify the acoustic cues responsible.

\paragraph{Exclusion of all-fail samples.}
Samples where all three budgets have zero \jf are termed \emph{all-fail pipeline cases}. The formal \cbl rule assigns them \textsc{Zero} solely through the lowest-cost tie-break, providing a cost-only fallback rather than a preference among successful budgets. The original supervision excludes these cases, but every downstream quality and cost evaluation retains them. There are 165 of 2,287 such cases on test\_s, 68 of 1,646 on test\_u, and 15 of 219 on the original R$^2$-AVSBench split. Their status does not identify the failing pipeline component.

\paragraph{All-fail supervision sensitivity.}
\label{app:allfail}
We compare E0, the original supervision excluding the 165 all-fail test\_s cases, with E1, which includes their cost-tie \textsc{Zero} labels. Both use the same 4,096-dimensional features, $4096\to256\to3$ MLP, ReLU, 0.1 dropout, Adam at $10^{-3}$, batch size 32, 50-epoch limit, early stopping, class-balanced loss, five out-of-fold splits, and seed 42. All-fail cases retain their assigned training, validation, and held-out membership; E0 excludes them only from training and validation supervision. E1 recomputes class weights within each training fold. No E1-specific retuning is performed, and both controllers are evaluated on all 2,287 raw samples.

\begin{table}[htbp]
\centering
\caption{All-fail supervision sensitivity. Classification metrics use the 2,122 originally informative cases; full-raw quality and cost use all 2,287 cases.}
\label{tab:allfail_sensitivity}
\begin{tabular}{lcc}
\toprule
Metric & E0: exclude & E1: include cost-tie ZERO \\
\midrule
Full-raw \jf & 0.5280 & 0.5278 \\
Mean reasoning tokens & 110.0 & 105.3 \\
Retention vs.\ Always-Long (0.549) & 96.2\% & 96.1\% \\
\midrule
Informative-case \jf & 0.5691 & 0.5688 \\
Three-class accuracy & 0.546 & 0.544 \\
Binary balanced accuracy & 0.663 & 0.660 \\
\midrule
All-fail ZERO rate & 41\% & 72\% \\
All-fail mean / median tokens & 116 / 93 & 51 / 0 \\
All-fail p95 tokens & 275 & 275 \\
\bottomrule
\end{tabular}
\end{table}

Using 1,000 paired bootstrap resamples over sample IDs, the E1--E0 full-raw \jf difference is $-0.0002$, 95\% CI $[-0.0018,0.0014]$; its lower limit exceeds the specified $-0.005$ non-inferiority margin. The token difference is $-4.7$, 95\% CI $[-5.6,-3.8]$. On informative cases, the differences are $-0.0003$ \jf $[-0.0019,0.0015]$, $-0.002$ three-class accuracy $[-0.008,0.004]$, and $-0.003$ binary balanced accuracy $[-0.010,0.004]$.

The all-fail cost reduction contributes $(165/2287)(116-51)=4.69$ tokens to the population mean, close to the observed 4.7-token reduction. Every evaluated action scores zero on these samples, so changing their route directly affects cost but not their \jf. The unchanged p95 also shows that E1 does not identify every failing case. We do not interpret their cost-tie labels as successful-budget routing accuracy.

\section{Additional Diagnostic Tables}
\label{app:diagnostic_tables}
\label{app:diagnostics}

\paragraph{Class-sensitive binary evaluation.}
The binary target distinguishes need-zero from need-reasoning, the union of need-short and need-long. Balanced accuracy is the unweighted mean of the two class recalls. The three-way router collapsed to binary and the separately trained binary MLP are distinct predictors; the latter obtains 0.689 accuracy. The same-dimensional Gaussian control remains near the reported majority result under the original protocol, so dimensionality alone does not explain the onset-feature gain.

\begin{table}[htbp]
\centering
\caption{Class-sensitive evaluation on the original informative test\_s population. Dashes indicate unreported individual recalls; exact metrics are not reconstructed from the rounded figure.}
\label{tab:balanced_metrics}
\begin{tabular}{lcccc}
\toprule
Readout & Accuracy & Balanced acc. & Zero recall & Reasoning recall \\
\midrule
Binary linear probe & 0.701 & 0.700 & 0.692 & 0.708 \\
Three-way router, collapsed & 0.665 & 0.663 & --- & --- \\
Reported majority reference & 0.577 & 0.500 & 0.000 & 1.000 \\
\bottomrule
\end{tabular}
\end{table}

\paragraph{Margin-stratified routing.}
Using the original out-of-fold controller predictions, we define
\begin{equation}
m(s)=\Phi^*(s)-\max_{b\ne b^*(s)}\Phi(s,b),
\end{equation}
where $b^*(s)$ is the cost-aware \cbl label. A cheaper within-tolerance action that is not a strict maximizer has $m(s)=0$. Splitting the 2,122 informative cases at the median gives Table~\ref{tab:margin_stratification}. Need-short errors concentrate near weakly separated action outcomes. This supports label-boundary ambiguity, but does not establish stochastic label noise or that every classification error incurs little downstream regret.

\begin{table}[htbp]
\centering
\caption{Margin-stratified performance. Binary accuracy collapses the original three-way controller's predictions.}
\label{tab:margin_stratification}
\begin{tabular}{lccc}
\toprule
Subset & Three-class accuracy & Binary accuracy & Need-short recall \\
\midrule
High-separation half & 0.641 & 0.752 & 0.542 \\
Near-tie half & 0.451 & 0.578 & 0.178 \\
\bottomrule
\end{tabular}
\end{table}

\begin{table}[h]
\centering\small
\caption{\textbf{Empirical quantification of the overthinking trap.} The audit uses 200 need-zero examples satisfying $\Phi(s,b_0)-\Phi(s,b_L)>\tau$. It characterizes this selected harmful-long subset, rather than the entire need-zero class or full test population.}
\label{tab:error_analysis}
\setlength{\tabcolsep}{5pt}
\begin{tabular}{lcc}
\toprule
\textbf{Metric} & \textbf{Original / \textsc{Zero}} & \textbf{\textsc{Long}-generated phrase} \\
\midrule
Avg. grounding phrase length (words) & 6.2 & 21.4 (\textcolor{red}{$\uparrow 245\%$}) \\
Unsupported additions & N/A & 41.5\% (\textcolor{red}{$\uparrow$}) \\
GroundingDINO top-1 box IoU & 0.72 & 0.58 (\textcolor{red}{$\downarrow 0.14$}) \\
Target \jf score & 0.586 & 0.421 (\textcolor{red}{$\downarrow 0.165$}) \\
\bottomrule
\end{tabular}
\end{table}

\begin{table}[h]
\centering\small
\caption{\textbf{Projection removal ablation with retrained probes.} Probe accuracy uses informative cases; downstream \jf uses the full raw population. The ablation concerns representations supplied to the external readout, not activation patching during MLLM generation.}
\label{tab:projection}
\setlength{\tabcolsep}{4pt}
\begin{tabular}{lccc}
\toprule
\textbf{Onset representation} & \textbf{Evaluation probe} & \textbf{Bin.\ acc.} & \textbf{Downstream \jf} \\
\midrule
Original \honset & Normally trained & 0.701 & 0.528 \\
\midrule
\honset $-$ random direction & Retrained probe & 0.695 (\textcolor{gray}{$\downarrow 0.006$}) & 0.525 \\
\honset $-$ budget direction & \textbf{Retrained probe} & \textbf{0.612} (\textcolor{red}{$\downarrow 0.089$}) & \textbf{0.480} \\
\bottomrule
\end{tabular}
\end{table}

\paragraph{Projection uncertainty.}
Under 1,000 paired bootstrap resamples, the retrained-probe accuracy change in Table~\ref{tab:projection} is $-0.089$, 95\% CI $[-0.103,-0.075]$. Routed \jf changes by $-0.048$ $[-0.056,-0.040]$, compared with $-0.003$ $[-0.008,0.002]$ for random-direction removal. These results support the selected direction's usefulness for the external readout/routing pipeline. They do not establish that the frozen MLLM uses the direction during its own generation.

\section{Hand-Crafted Signal Results}
\label{app:signal_table}

Table~\ref{tab:signals} reports the original proxy features and their readouts. The four-signal MLP obtains 0.541 binary accuracy versus 0.689 for the onset-state MLP. This comparison concerns the evaluated summaries, rather than establishing irreducibility or a distributed causal mechanism. The scalar AUCs retain their reported orientation; reversing a feature can reverse its AUC. The learned proxy classifiers provide the corresponding supervised comparison.

\begin{table}[h]
\caption{\textbf{Original interpretable-proxy results.} Individual rows use binary AUC; joint classifiers use binary accuracy. The four-signal MLP is 14.8 points below the onset MLP. Joint classifiers use the same class-balanced training protocol as the controller and are not optimized to reproduce the raw majority-class baseline.}
\label{tab:signals}
\centering\small
\setlength{\tabcolsep}{4pt}
\begin{tabular}{@{}llcc@{}}
\toprule
Input (dim) & Metric & Score & 95\% CI / Note \\
\midrule
Attn.\ entropy $E_{vt}$ (1) & Bin.\ AUC & 0.497 & [0.474, 0.521] \\
Attn.\ entropy $E_{at}$ (1) & Bin.\ AUC & 0.508 & [0.485, 0.531] \\
Latent shift $\delta$ (1)   & Bin.\ AUC & 0.495 & [0.472, 0.519] \\
First-tok.\ ent.\ $H_1$ (1) & Bin.\ AUC & 0.466 & [0.442, 0.491] \\
\midrule
4 signals, logistic (4) & Bin.\ Accuracy & 0.526 & $\Delta{=}{-}0.163$ vs.\ \honset \\
4 signals, MLP (4)      & Bin.\ Accuracy & 0.541 & $\Delta{=}{-}0.148$ vs.\ \honset \\
\textbf{\honset MLP (ref.)} & Bin.\ Accuracy & \textbf{0.689} & --- \\
\bottomrule
\end{tabular}
\end{table}
\paragraph{Stronger proxy features.}
The additional features in Table~\ref{tab:stronger_proxies} are evaluated under the same fixed five-fold, held-out, class-balanced MLP protocol. The combined feature set narrows the gap to the onset MLP from 14.8 to 10.3 percentage points. The supported conclusion is that the evaluated summaries do not recover all predictive information available to the onset readout; other feature families, including learned sparse dictionaries, remain untested.

\begin{table}[htbp]
\centering
\caption{Stronger interpretable-proxy comparison. Each addition is made to the four original signals; all rows use an MLP. Gaps are relative to onset accuracy 0.689.}
\label{tab:stronger_proxies}
\begin{tabular}{lcc}
\toprule
Input & Binary accuracy & Gap to onset MLP \\
\midrule
Four signals & 0.541 & $-0.148$ \\
Four signals + first-token logit margin & 0.552 & $-0.137$ \\
Four signals + per-layer entropy profile & 0.578 & $-0.111$ \\
All combined & 0.586 & $-0.103$ \\
Onset state & 0.689 & --- \\
\bottomrule
\end{tabular}
\end{table}

\section{Computational Cost Analysis}
\label{app:cost}

Table~\ref{tab:cost} reports the original rounded model-stage costs. The onset controller adds little routing overhead while reducing average generation cost in the evaluated configuration. These measurements concern reasoning tokens and runtime on the stated hardware; they do not establish energy, memory, or throughput savings in other serving configurations. The direct paired latency benchmark below distinguishes the time of the routing decision from completion of the MLLM and of the whole segmentation pipeline.

\begin{table}[h]
\caption{\textbf{Original model-stage cost comparison on RefAVSBench test\_s.} Routing denotes budget-dispatch overhead. MLLM latency includes encoding, routing, and generation, but excludes the subsequent detector and segmenter. Timings are rounded averages on a single NVIDIA A100 80GB.}
\label{tab:cost}
\centering\small
\setlength{\tabcolsep}{5pt}
\begin{tabular}{@{}lcccc@{}}
\toprule
Method & \#Params (route) & Avg tok & Route time & MLLM time \\
\midrule
always-zero                      & ---    &   0 & ---      & 0.8\,s \\
always-short                     & ---    &  93 & ---      & 2.1\,s \\
always-long (TGS-Agent)          & ---    & 275 & ---      & 4.7\,s \\
\midrule
\textbf{Ours} (adaptive \honset) & 1.1\,M & 110 & $<$1\,ms & 2.3\,s \\
\bottomrule
\end{tabular}
\end{table}

\paragraph{Offline cost and amortization.}
\cbl requires native action outcomes and target masks to compute segmentation utility. The original study-wide total of 85 A100-hours includes analyses beyond training one deployed controller. Charging all of it against the rounded $4.7-2.3=2.4$\,s/query MLLM saving gives approximately 127,500 queries ($85\times3600/2.4$). A narrower MLLM-only calculation costs $2287(0.8+2.1+4.7)=17{,}381.2$\,s, approximately 4.83 A100-hours, corresponding to about 7,242 queries. All raw samples must be evaluated before all-fail cases can be identified. This lower accounting boundary excludes detector/segmenter evaluation and other setup work. Across all 4,152 original raw samples, its corresponding model component is approximately 8.8 A100-hours. The historical 7.2k--127.5k range is an accounting bracket, not a universal deployment guarantee. Appendix~\ref{app:calibration} reports later complete setup measurements and explicitly separates full-pool ratios from fold-matched break-even calculations.

\subsection{Direct paired latency benchmark}
\label{app:latency}
All 2,287 raw test\_s samples, including all-fail cases, are evaluated on one A100 80GB at batch size one. We hold weights, prompts, budget definitions, out-of-fold controllers, greedy decoding, stopping criteria, precision, and downstream settings fixed. Timing begins from the same model-ready input. Time-to-decision (TTD) ends when the final budget is selected; MLLM latency ends when the model output is complete, before grounding and segmentation. Short-then-Decide generates SHORT before its secondary decision and LONG upgrade. The onset router encodes once, reads \honset, and generates only the selected action.

We use 50 warm-ups, a fixed seeded sample order, AB/BA counterbalancing, three repetitions per sample--method pair, and CUDA synchronization. Each pair is summarized by its median; no outlier is removed. Confidence intervals use 1,000 paired bootstrap resamples over sample IDs. Repetitions are not treated as independent samples.

\begin{table}[htbp]
\centering
\caption{Paired benchmark of the original controllers. Timing triplets are mean / median / p95 in seconds. MLLM time excludes the downstream perception stack.}
\label{tab:paired_latency_appendix}
\begin{tabular}{lcccc}
\toprule
Method & \jf & Tokens & TTD (s) & MLLM time (s) \\
\midrule
Short-then-Decide & 0.535 & 182 & 2.13 / 2.07 / 2.61 & 3.17 / 3.02 / 5.86 \\
Onset router & 0.528 & 110 & 0.82 / 0.80 / 0.96 & 2.31 / 2.18 / 4.75 \\
\bottomrule
\end{tabular}
\end{table}

The paired onset-minus-Short-then-Decide differences are $-1.31$\,s TTD, 95\% CI $[-1.34,-1.28]$, and $-0.86$\,s MLLM latency $[-0.94,-0.78]$. The onset decision occurs first on 99.8\% of samples. Mean decision time and model latency are reduced by 61.5\% and 27.1\%, respectively. The token difference is $-72$ $[-75,-69]$, or 39.6\% fewer reasoning tokens, while \jf is 0.007 lower. The original controller therefore retains 98.7\% of Short-then-Decide's quality at a lower measured latency and cost point.

As a secondary implementation sensitivity, semantically admissible cache reuse on the Short-then-Decide upgrade path reduces its mean MLLM latency to 2.92\,s. The onset router remains at 2.31\,s, a paired difference of $-0.61$\,s $[-0.69,-0.53]$. TTD is unchanged because the SHORT prefix must finish before the secondary decision. Descriptive complete-pipeline mean/p95 latencies are 7.28/10.10\,s for Short-then-Decide, 7.02/9.42\,s with cache reuse, and 6.42/8.96\,s for onset routing. No additional paired significance claim is attached to those end-to-end aggregates. The findings concern the tested batch-one hardware configuration.

\section{Additional Figures}
\label{app:additional_figs}

The following figures supplement the quantitative analysis with proxy-feature diagnostics, qualitative examples, quality--cost operating points, and per-label routing behavior.

\begin{figure}[htbp]
\centering
\includegraphics[width=\linewidth]{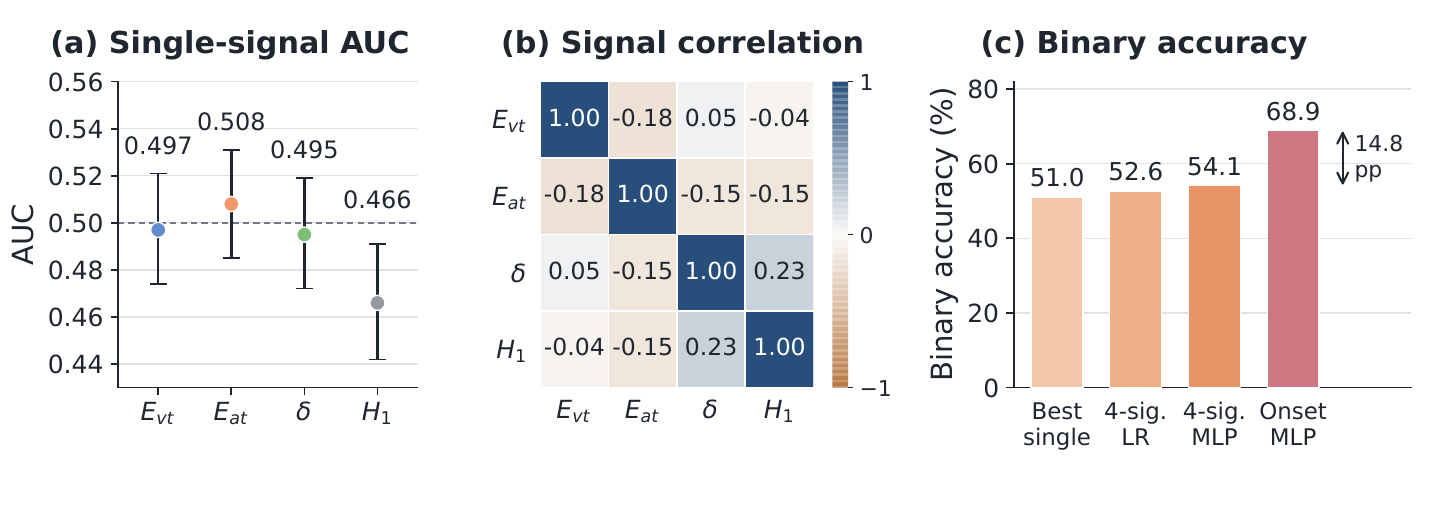}
\caption{\textbf{Original proxy-feature diagnostics.} (a) Individual scalar proxy AUCs with 95\% confidence intervals. (b) Pairwise proxy correlations. (c) Original binary-accuracy comparison. The four-signal MLP has a 14.8-percentage-point accuracy gap to the onset MLP; stronger evaluated proxies reduce the gap to 10.3 points (Table~\ref{tab:stronger_proxies}). These comparisons do not establish a distributed causal mechanism.}
\label{fig:distributed}
\end{figure}

\begin{figure}[htbp]
\centering
\includegraphics[width=0.8\linewidth]{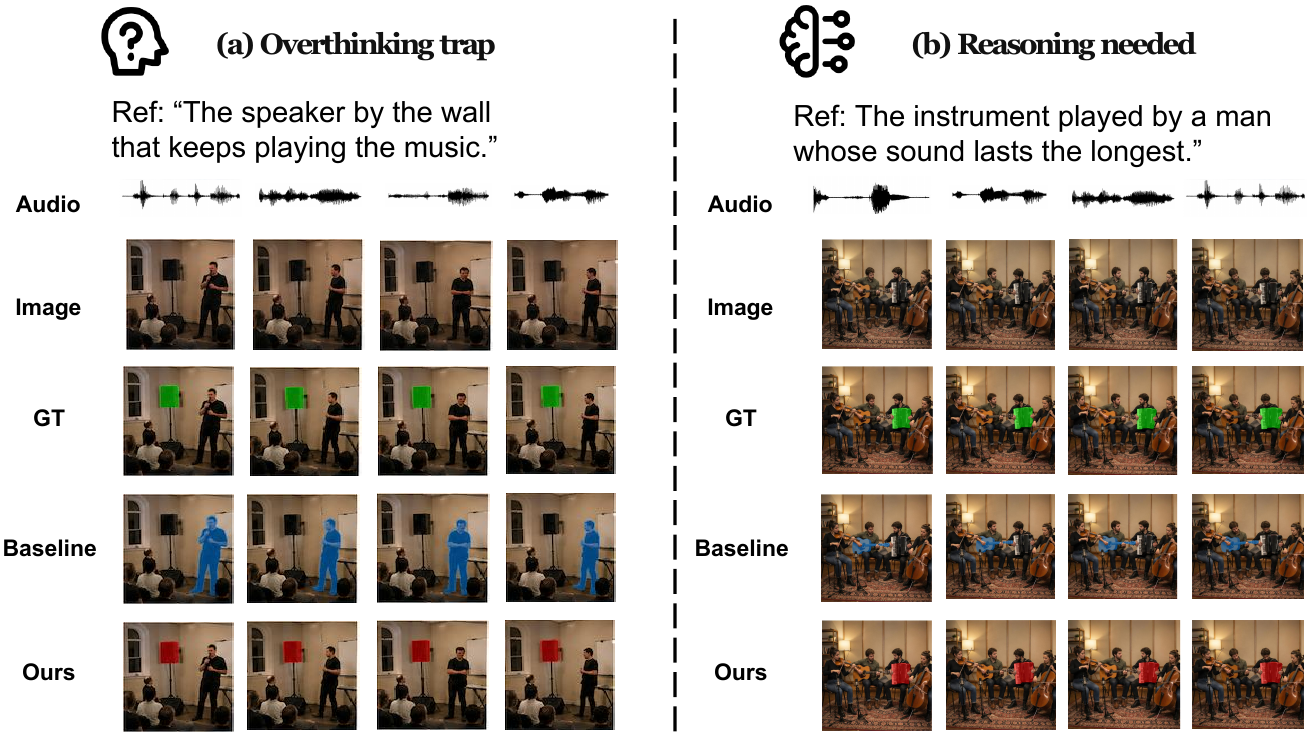}
\caption{\textbf{Qualitative comparison of adaptive reasoning budgets.} (a) \textbf{Overthinking trap:} fixed long reasoning drifts to the wrong subject, while our zero-budget route preserves visual focus. (b) \textbf{Reasoning needed:} our long-budget route resolves the ambiguous instrument. GT masks are green.}
\label{fig:qualitative_main}
\end{figure}

\begin{figure}[htbp]
\centering
\includegraphics[width=0.78\linewidth]{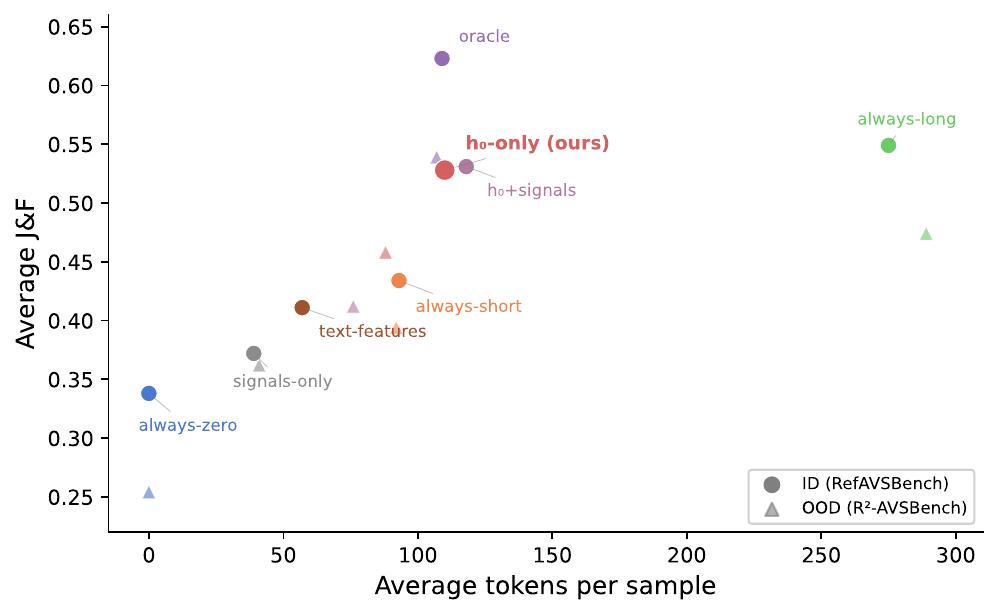}
\caption{\textbf{Quality--cost operating points on RefAVSBench test\_s.} The onset-only controller obtains approximately 96\% of Always-Long \jf at 40\% of its reasoning-token cost. The outcome-informed \cbl point is a cost-aware policy reference, not a strict maximum-quality upper bound. R$^2$ markers describe transfer without target-label fitting; the transferred router's 0.458 \jf remains below Always-Long's 0.474.}
\label{fig:pareto}
\end{figure}

\FloatBarrier
\begin{figure}[H]
\centering
\includegraphics[width=\linewidth]{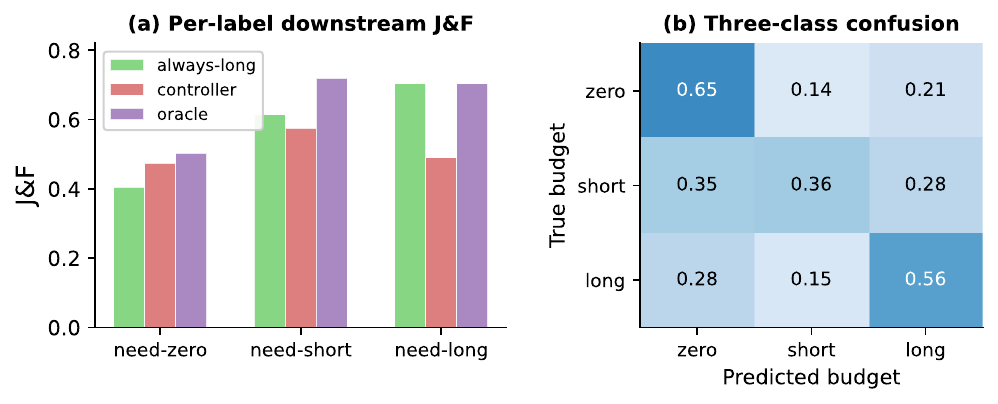}
\caption{\textbf{Per-label controller analysis.} (a) Downstream \jf grouped by the cost-aware \cbl label. (b) Three-class confusion matrix of the onset-only controller; rows are rounded. Need-short has the lowest recall. The additional margin stratification (Table~\ref{tab:margin_stratification}) shows that its errors are concentrated among weakly separated budget outcomes, rather than establishing an intrinsically continuous transition class.}
\label{fig:perlabel}
\end{figure}

\section{Native Calibration and Deployment Cost}
\label{app:calibration}

\subsection{Calibration protocol and source-domain sensitivity}

The dependence of \cbl on a fixed pipeline motivates a direct test of the amount of native supervision required after a change. Native calibration means that the candidate reasoning policies are executed with the target pipeline and scored against target masks. Calibration samples are selected from raw training IDs before their native labels are inspected. Every raw held-out case, including cases on which all candidate actions fail, remains in the evaluation population. Thus, the calibration fraction controls the amount of acquired supervision rather than the difficulty of the test population.

Reduced-calibration conditions are compared with matched full-calibration controllers through out-of-fold evaluation. The original three-action controller at 0.528 \jf and 110 tokens is retained as a separate reference; it is not substituted for a retrained full-data control. Differences are defined as reduced minus matched full calibration. The quality criterion requires the lower endpoint of the 95\% interval on $\Delta\jf$ to exceed $-0.005$, while the cost criterion requires the upper endpoint on $\Delta$tokens to remain below $+5$. The reported principal conditions satisfy both criteria and the corresponding seed-level criteria in all five seeds. Native masks and counterfactual outcomes are still required for the selected subset.

\begin{table}[htbp]
\centering
\caption{Source-domain calibration sensitivity. Each fraction is applied to the raw training pool within a fold. Scores are full-raw held-out results; uncertainties are standard deviations across the matched runs. Informative-label counts exclude all-fail training cases. The separately reproduced original controller obtains 0.528 \jf at 110 tokens.}
\label{tab:source_calibration_curve}
\begin{tabular}{rrrrr}
\toprule
Fraction & Raw IDs/fold & Informative/fold & \jf & Tokens \\
\midrule
5\% & 92 & $85.4\pm1.6$ & $0.513\pm0.004$ & $114\pm3$ \\
10\% & 183 & $169.8\pm2.3$ & $0.520\pm0.003$ & $112\pm2$ \\
25\% & 458 & $424.7\pm3.5$ & $0.527\pm0.002$ & $110\pm2$ \\
50\% & 915 & $849.1\pm4.8$ & $0.529\pm0.002$ & $109\pm1$ \\
100\% & 1,829--1,830 & $1,697.6\pm0.5$ & $0.530\pm0.002$ & $109\pm1$ \\
\bottomrule
\end{tabular}
\end{table}

One-quarter of the source calibration pool is sufficient to preserve the retrained full-data operating point within the specified margins. At 25\%, the quality difference is $-0.003$ with interval $[-0.0044,-0.0016]$, and the token difference is $+1$ with interval $[-0.6,2.6]$. The smaller 5\% and 10\% conditions have quality differences of $-0.017$ and $-0.010$, respectively, and fall outside the quality margin. This curve locates an empirical calibration regime rather than implying that arbitrarily small label sets are adequate.

\subsection{Group-disjoint distribution adaptation}

Native adaptation on R$^2$-AVSBench is separated from the source-only transfer experiment. Because the benchmark rewrites references over RefAVS videos, the primary source initialization excludes every source-video cluster represented in the target benchmark. This group-disjoint source controller is denoted R0-D; Rp denotes its refresh with $p$ percent native calibration. The historical source-only controller, R0-S, and the same-video sensitivity, R0-M, are reported separately. The full raw target population contains 219 examples in each condition.

\begin{table}[htbp]
\centering
\caption{Native R$^2$ calibration. R0-D, Rp, and R100 form the primary group-disjoint comparison. The historical R0-S anchor and the same-video (M) and head-only (H) variants are separate sensitivities. Counts denote informative training labels per fold; all 219 raw target cases remain in evaluation.}
\label{tab:r2_calibration_curve}
\begin{tabular}{lrrr}
\toprule
Condition & Native labels/fold & \jf & Tokens \\
\midrule
R0-S, historical & 0 & 0.458 & 112 \\
R0-M, same-video & 0 & $0.463\pm0.003$ & $111\pm3$ \\
R0-D, group-disjoint & 0 & $0.452\pm0.004$ & $114\pm4$ \\
R5 & $8.4\pm0.5$ & $0.460\pm0.005$ & $112\pm4$ \\
R10 & $16.7\pm0.7$ & $0.466\pm0.004$ & $110\pm3$ \\
R20 & $32.6\pm1.1$ & $0.470\pm0.003$ & $108\pm2$ \\
R50 & $81.9\pm1.4$ & $0.472\pm0.003$ & $107\pm2$ \\
R100 & 163.2 & $0.471\pm0.003$ & $108\pm2$ \\
\midrule
R20-M & $32.6\pm1.1$ & $0.471\pm0.003$ & $107\pm2$ \\
R100-M & 163.2 & $0.472\pm0.003$ & $107\pm2$ \\
R20-H & $32.6\pm1.1$ & $0.466\pm0.004$ & $111\pm3$ \\
R100-H & 163.2 & $0.468\pm0.004$ & $110\pm3$ \\
\bottomrule
\end{tabular}
\end{table}

Most of the full-refresh gain is recovered by 20\% native calibration. R100 improves on R0-D by $0.019$ \jf with interval $[0.013,0.025]$. The R20 difference from R100 is $-0.001$ with interval $[-0.0032,0.0012]$, while its token difference is zero with interval $[-1.8,1.8]$. The 90\%-recovery contrast is positive, at $0.0009$ with interval $[0.0002,0.0016]$. At the reported operating points, R20 also exceeds Short-then-Decide by $0.009$ \jf and uses 75 fewer tokens. These are supervised adaptation results with a group-disjoint source initialization; the source-only 0.458 result retains its original transfer interpretation.

\begin{table}[htbp]
\centering
\caption{Paired quality and cost differences for selected native-calibration conditions. Each entry is reduced calibration minus its own full-calibration reference, with a 95\% interval. The quality and token margins are $-0.005$ and $+5$, respectively.}
\label{tab:calibration_paired_intervals}
\begin{tabular}{lrr}
\toprule
Comparison & $\Delta\jf$ & $\Delta$tokens \\
\midrule
RefAVS 25\% $-$ 100\% & $-0.003\;[-0.0044,-0.0016]$ & $+1\;[-0.6,2.6]$ \\
R20 $-$ R100 & $-0.001\;[-0.0032,0.0012]$ & $0\;[-1.8,1.8]$ \\
Q20-N $-$ Q100-N & $-0.002\;[-0.0043,0.0003]$ & $+1\;[-0.8,2.8]$ \\
Q20-R $-$ Q100-R & $-0.001\;[-0.0030,0.0010]$ & $+1\;[-0.8,2.8]$ \\
\bottomrule
\end{tabular}
\end{table}

\subsection{Native backbone calibration and label accounting}

Native Qwen calibration evaluates a different question from the transferred-label probe. The earlier binary accuracy of 0.704 measures concordance with labels defined by the primary pipeline. Here, target-native outcomes are generated and native controllers are evaluated through downstream segmentation. Two supervision treatments are distinguished: Qp-R replaces selected source labels with native labels while retaining the remaining transferred table, whereas Qp-N uses only the selected native supervision. In the latter condition, no transferred Llama label is used.

\begin{table}[htbp]
\centering
\caption{Native Qwen refresh. R denotes replacement in a retained transferred-label pool; N denotes native-only supervision. Each reduced condition is compared with the full reference from the same family. Q0-Transfer is evaluated through segmentation, separately from the transferred-label probe.}
\label{tab:qwen_calibration_curve}
\begin{tabular}{lrrr}
\toprule
Condition & Native labels/fold & \jf & Tokens \\
\midrule
Q0-Transfer & 0 & $0.512\pm0.004$ & $119\pm4$ \\
Q5-R & $86.3\pm2.0$ & $0.522\pm0.004$ & $115\pm4$ \\
Q10-R & $171.7\pm2.8$ & $0.532\pm0.003$ & $110\pm3$ \\
Q20-R & $343.1\pm4.2$ & $0.538\pm0.003$ & $106\pm2$ \\
Q50-R & $857.6\pm5.1$ & $0.540\pm0.002$ & $105\pm2$ \\
Q100-R & $1,716.0\pm0.8$ & $0.539\pm0.002$ & $105\pm2$ \\
\midrule
Q20-N & $343.1\pm4.2$ & $0.536\pm0.003$ & $107\pm2$ \\
Q100-N & $1,716.0\pm0.8$ & $0.538\pm0.002$ & $106\pm2$ \\
\bottomrule
\end{tabular}
\end{table}

The native-only condition preserves the tested full-native point without requiring a source table. Q20-N obtains 0.536 \jf at 107 tokens, compared with 0.538 at 106 tokens for Q100-N. Both non-inferiority criteria are satisfied. In the replacement family, Q20-R recovers more than 90\% of the gain from Q0-Transfer to Q100-R, with recovery contrast $0.0017$ and interval $[0.0006,0.0028]$. The native Qwen Always-Short reference obtains 0.535 \jf at 91 tokens, exceeding the transferred-label router's quality while using fewer tokens. Native supervision therefore matters for downstream routing, although this comparison does not identify the direction of bias in the separate 0.704 concordance measurement.

All-fail status changes with the backbone as well as with the action space. Of the 2,287 raw source examples, 2,145 are informative under native Qwen outcomes and 142 are all-fail. Among the 165 original source all-fail examples, 140 become informative for Qwen. Native labels therefore cannot be obtained by simply preserving the source informative subset. The full-pool counts in Table~\ref{tab:qwen_label_accounting} document this change; they describe a one-controller deployment analogue rather than per-fold training counts.

\begin{table}[htbp]
\centering
\caption{Full-pool native-Qwen label accounting. ``Removed'' denotes selected native all-fail cases omitted from supervision, not from evaluation. Counts in the upper panel are supervised rows; the lower panel shows the overlap of informative populations across the two pipelines.}
\label{tab:qwen_label_accounting}
\begin{tabular}{lrrrr}
\toprule
Condition & Transferred & Native & Removed & Total \\
\midrule
Q0-Transfer & 2,122 & 0 & 0 & 2,122 \\
Q20-R & 1,698 & 430 & 28 & 2,128 \\
Q100-R & 0 & 2,145 & 142 & 2,145 \\
Q20-N & 0 & 430 & 28 & 430 \\
Q100-N & 0 & 2,145 & 142 & 2,145 \\
\bottomrule
\end{tabular}

\medskip
\begin{tabular}{lrrr}
\toprule
Source population & Qwen informative & Qwen all-fail & Total \\
\midrule
Informative & 2,005 & 117 & 2,122 \\
All-fail & 140 & 25 & 165 \\
Total & 2,145 & 142 & 2,287 \\
\bottomrule
\end{tabular}
\end{table}

\subsection{Complete setup and conditional amortization}

Reduced calibration lowers measured setup under the tested deployment assumptions. Complete setup includes trajectory generation, detector and segmenter execution, onset extraction, controller training, serialization, startup, and other measured overhead on one A100. The full-pool measurements correspond to a single controller trained on the indicated raw calibration pool. Stage-specific GPU-hours and wall-clock elapsed seconds are reported separately because they describe different aspects of resource use.

\begin{table}[htbp]
\centering
\caption{Complete full-pool setup measurements. The three stage columns are GPU-hours for the MLLM, detector/segmenter (D/S), and controller training. Elapsed time additionally includes extraction, startup, serialization, and overhead. R$^2$ costs are incremental given the same group-disjoint source controller.}
\label{tab:complete_setup_components}
\begin{tabular}{lrrrrr}
\toprule
Condition & Raw IDs & MLLM & D/S & Train & Seconds \\
\midrule
RefAVS 100\% & 2,287 & 4.83 & 1.36 & 0.05 & 22,610 \\
RefAVS 25\% & 572 & 1.21 & 0.34 & 0.04 & 5,790 \\
R100 & 219 & 0.462 & 0.130 & 0.022 & 2,210 \\
R20 & 44 & 0.093 & 0.031 & 0.019 & 515 \\
Q100-R & 2,287 & 4.59 & 1.36 & 0.05 & 21,600 \\
Q20-R, table sunk & 458 & 1.26 & 0.34 & 0.05 & 5,960 \\
Q20-R, cold start & 2,287 & 5.21 & 1.36 & 0.06 & 24,140 \\
Q100-N & 2,287 & 4.59 & 1.36 & 0.05 & 21,600 \\
Q20-N & 458 & 0.92 & 0.34 & 0.05 & 4,750 \\
\bottomrule
\end{tabular}
\end{table}

The calibration fraction must be interpreted against the appropriate denominator. For example, 25\% RefAVS calibration uses 458 raw training examples within a held-out fold, whereas the full-pool analogue uses 572 of 2,287 examples. Generated-token counts for the full/25\% RefAVS analogues are 841,616/210,496; R100/R20 use 80,592/16,192. Q100-N/Q20-N use 821,033/164,422. These counts quantify acquired counterfactual trajectories rather than tokens incurred on a future routed query.

\begin{table}[htbp]
\centering
\caption{Complete setup ratios against matched full calibration. Intervals are the reported 95\% intervals. The cold-start replacement condition charges construction of the retained source table and is more expensive than full native calibration; the native-only condition requires no such table.}
\label{tab:complete_setup_ratios}
\begin{tabular}{lr}
\toprule
Reduced condition / matched full reference & Setup ratio \\
\midrule
RefAVS 25\% / 100\% & $0.256\;[0.244,0.269]$ \\
R20 / R100 & $0.233\;[0.209,0.258]$ \\
Q20-N / Q100-N & $0.220\;[0.205,0.236]$ \\
Q20-R / Q100-R, table sunk & $0.276\;[0.258,0.296]$ \\
Q20-R / Q100-R, cold start & $1.118\;[1.074,1.163]$ \\
\bottomrule
\end{tabular}
\end{table}

Break-even is computed within a matched deployment population. For a fold-specific setup cost $C_f$ and held-out mean online saving $\Delta t_f>0$, the corresponding query count is $Q_{\mathrm{BE},f}=C_f/\Delta t_f$. The reported upper 95\% estimates use the setup and latency of the corresponding out-of-fold controller. They are not obtained by dividing the preceding full-pool elapsed times by an aggregate online saving. The distinction prevents the cost of one calibration population from being combined with savings measured for another controller.

\begin{table}[htbp]
\centering
\caption{Fold-matched amortization against Always-Long. Online savings are seconds per held-out query. The last column gives the reported upper 95\% break-even estimate for each condition. These estimates and the full-pool setup measurements have intentionally different populations.}
\label{tab:foldmatched_break_even}
\begin{tabular}{lrr}
\toprule
Condition & Saving/query (s) & Upper break-even (queries) \\
\midrule
RefAVS 100\% & 2.39 & 8,910 \\
RefAVS 25\% & 2.38 & 2,310 \\
R100 & 2.39 & 1,130 \\
R20 & 2.40 & 245 \\
Q100-N & 2.37 & 10,180 \\
Q20-N & 2.36 & 2,070 \\
Q100-R & 2.38 & 10,150 \\
Q20-R, table sunk & 2.37 & 2,270 \\
Q20-R, cold start & 2.37 & 9,980 \\
\bottomrule
\end{tabular}
\end{table}

Historical supervision materially changes the accounting. With an existing source table treated as sunk, Q20-R costs 5,960 seconds; charging its construction increases the same controller's setup to 24,140 seconds, exceeding the 21,600-second full-native reference. Accordingly, the replacement result supports inexpensive refresh only when the retained table genuinely predates migration. Q20-N supports a separate native-only scenario and reduces complete measured setup to 22.0\% of its full-native reference.

The earlier 7.2k--127.5k query bracket used MLLM-only and study-wide accounting, respectively, and should be distinguished from these measured deployment scenarios. The supported regime remains a fixed or slowly evolving pipeline serving enough requests to amortize calibration. Native relabeling is reduced for the tested changes, while abrupt prompt, perception-stack, action-set, or domain changes can require a new calibration study.

\FloatBarrier
\section{Finer Action Spaces and Bounded Stage Exits}
\label{app:actions}

\subsection{Semantic policies and action-dependent populations}

The three original actions provide a tractable counterfactual design, but their granularity can limit the achievable frontier. Four intermediate semantic-depth policies are therefore added between SHORT and LONG. Progressively deeper audio, visual, temporal, and cross-modal resolution is requested while the perception stack remains fixed. The original set is $\mathcal B_3=\{b_0,b_1,b_6\}$, the cost-spaced subset is $\mathcal B_5=\{b_0,b_1,b_3,b_5,b_6\}$, and $\mathcal B_7$ includes all seven policies. The two additions in $\mathcal B_5$ are selected by cost spacing rather than segmentation quality.

\begin{table}[htbp]
\centering
\caption{Realized costs and validity of the semantic policies. Costs are median reasoning-token counts, not hard token quotas. Parse validity and hard cutoffs are measured on all 2,287 raw RefAVS cases. Crossing is the reported adjacent-policy token-order crossing rate.}
\label{tab:semantic_action_validity}
\begin{tabular}{lrrrr}
\toprule
Action & Median tokens & Parse-valid & Hard cutoff & Crossing \\
\midrule
$b_0$, ZERO & 0 & 100.0\% & 0.0\% & --- \\
$b_1$, SHORT & 93 & 99.9\% & 0.1\% & 7.4\% \\
$b_2$ & 124 & 99.8\% & 0.2\% & 11.6\% \\
$b_3$ & 154 & 99.8\% & 0.2\% & 13.2\% \\
$b_4$ & 187 & 99.7\% & 0.3\% & 15.1\% \\
$b_5$ & 229 & 99.7\% & 0.3\% & 17.6\% \\
$b_6$, LONG & 275 & 99.6\% & 0.4\% & --- \\
\bottomrule
\end{tabular}
\end{table}

The new policies supply usable intermediate operating points with distinct realized costs. The maximum cutoff incidence is 0.4\%, while parse validity remains at least 99.6\%. The median lengths should nevertheless not be interpreted as a strict per-example ordering: content-dependent generation produces adjacent token crossings. Counterfactual comparisons therefore use the realized action outcomes and costs rather than assuming that every longer named policy always emits more tokens.

\begin{table}[htbp]
\centering
\caption{Action-dependent informative populations and tolerance-based \cbl policies. Every row is evaluated on the same 2,287 raw cases. All-fail denotes zero \jf under every action in the indicated set. An oracle action is the least costly candidate within tolerance $\tau$ of the best sample-level quality.}
\label{tab:semantic_oracle_populations}
\begin{tabular}{lrrrr}
\toprule
Action set & Informative & All-fail & Policy \jf & Mean tokens \\
\midrule
$\mathcal B_3$ & 2,122 & 165 & 0.623 & 109 \\
$\mathcal B_5$ & 2,133 & 154 & 0.633 & 106 \\
$\mathcal B_7$ & 2,138 & 149 & 0.638 & 104 \\
\bottomrule
\end{tabular}
\end{table}

Finer policies improve the tolerance-based oracle by 0.015 \jf while reducing its average cost by five tokens. Of this increase, 0.0141 is contributed by the original 2,122 informative examples, and 0.0009 by original all-fail examples rescued by an intermediate action. Thus, the improvement is concentrated in the original evaluation population rather than being created by changing its denominator. The informative-case label distributions are 42.4/27.3/30.3\% for $\mathcal B_3$, 41.1/20.8/14.6/10.9/12.6\% for $\mathcal B_5$, and 40.7/18.2/10.1/8.4/7.8/6.9/7.9\% for $\mathcal B_7$, in increasing action-index order.

\subsection{Population quality--cost frontiers}

The tolerance-based \cbl policy and the population frontier answer different questions. The former imposes a sample-wise tolerance; the latter maximizes mean quality subject to an aggregate budget. The exact deterministic frontier selects one evaluated action per sample through mixed-integer optimization. Its linear-program relaxation permits convex mixtures and supplies an upper frontier. Both are outcome-aware diagnostic quantities rather than deployable routers.

\begin{table}[htbp]
\centering
\caption{Exact and relaxed population frontiers at 110 average reasoning tokens. The linear-program frontier exceeds its exact deterministic counterpart by 0.0003 for each action set. Both formulations preserve the finer-action advantage at this budget.}
\label{tab:semantic_exact_frontiers}
\begin{tabular}{lrrr}
\toprule
Action set & Exact \jf & Relaxed \jf & Relaxation gap \\
\midrule
$\mathcal B_3$ & 0.6287 & 0.6290 & 0.0003 \\
$\mathcal B_5$ & 0.6389 & 0.6392 & 0.0003 \\
$\mathcal B_7$ & 0.6442 & 0.6445 & 0.0003 \\
\bottomrule
\end{tabular}
\end{table}

\begin{table}[htbp]
\centering
\caption{Differences between relaxed frontiers with reported 95\% intervals. The first two columns evaluate fixed mean token budgets; the final column averages the gain over the stated 80--275-token range. The exact deterministic gain at 110 tokens agrees with each corresponding point estimate in the first column.}
\label{tab:semantic_frontier_gains}
\begin{tabular}{lrrr}
\toprule
Contrast & 110 tokens & 182 tokens & Mean, 80--275 \\
\midrule
$\mathcal B_5-\mathcal B_3$ & $0.0102$ & $0.0078$ & $0.0070$ \\
95\% interval & $[0.0083,0.0121]$ & $[0.0062,0.0094]$ & $[0.0057,0.0083]$ \\
$\mathcal B_7-\mathcal B_5$ & $0.0053$ & $0.0037$ & $0.0034$ \\
95\% interval & $[0.0038,0.0068]$ & $[0.0025,0.0049]$ & $[0.0024,0.0044]$ \\
$\mathcal B_7-\mathcal B_3$ & $0.0155$ & $0.0115$ & $0.0104$ \\
95\% interval & $[0.0132,0.0178]$ & $[0.0095,0.0135]$ & $[0.0087,0.0121]$ \\
\bottomrule
\end{tabular}
\end{table}

The seven-action advantage persists across operating budgets. At 110 tokens, the relaxed and exact gains over three actions are both 0.0155 \jf; at 182 tokens the relaxed gain is 0.0115. Averaging over the wider 0--275-token interval gives 0.0091 with interval $[0.0076,0.0106]$. The corresponding five-minus-three and seven-minus-five averages are 0.0061 and 0.0030. Hence the improvement is not confined to the original router's selected token point.

Returns diminish as the policy grid becomes denser. At 110 tokens, the five-action expansion contributes 0.0102 \jf, followed by a further 0.0053 from seven actions. Writing $F_K$ for the frontier at that budget, the second difference $(F_7-F_5)-(F_5-F_3)$ is $-0.0049$ with interval $[-0.0064,-0.0034]$. This indicates diminishing incremental gains in the tested semantic family, while the positive seven-minus-three difference rules out saturation at the original three actions.

\begin{table}[htbp]
\centering
\caption{Relaxed-frontier gains under matched three-action latency budgets, with reported 95\% intervals. The advantage is observed using both MLLM time and complete pipeline time as the resource constraint, rather than only reasoning-token counts.}
\label{tab:semantic_latency_frontiers}
\begin{tabular}{lrr}
\toprule
Contrast & Matched MLLM latency & Matched full latency \\
\midrule
$\mathcal B_5-\mathcal B_3$ & $0.0096\;[0.0076,0.0116]$ & $0.0089\;[0.0069,0.0109]$ \\
$\mathcal B_7-\mathcal B_5$ & $0.0050\;[0.0034,0.0066]$ & $0.0045\;[0.0030,0.0060]$ \\
$\mathcal B_7-\mathcal B_3$ & $0.0146\;[0.0121,0.0171]$ & $0.0134\;[0.0110,0.0158]$ \\
\bottomrule
\end{tabular}
\end{table}

The same headroom is observed when latency replaces tokens as the constraint. At the matched three-action MLLM and full-pipeline budgets, the seven-action gains are 0.0146 and 0.0134 \jf, respectively. Conversely, the relaxed minimum token costs to reach qualities 0.528, 0.535, and 0.549 are 56/63/76 for $\mathcal B_3$, 53/59/70 for $\mathcal B_5$, and 51/56/66 for $\mathcal B_7$. These outcome-aware frontiers motivate the practical question of whether an onset-only controller can exploit the added policies.

\subsection{Matched controllers and action masking}

Learned extensions are evaluated with matched onset features, shared trunks, folds, seeds, losses, and policy-selection procedures within each controller family. The action set and corresponding output heads are varied. The $\tau$-\cbl family predicts tolerance-based action labels. A second matched family is denoted Utility-$K$, with $K$ indicating the number of available actions. Its three-action control is retrained at 0.531 \jf and 110 tokens; this reference remains distinct from the original three-action controller at 0.528 and 110 tokens.

\begin{table}[htbp]
\centering
\caption{Matched learned controllers at full and 25\% calibration. Values are means and standard deviations across the matched runs. The original controller is reproduced separately and has no matched reduced-calibration counterpart in this table. Comparisons of action counts are made within each controller family.}
\label{tab:semantic_learned_controllers}
\begin{tabular}{lrrrr}
\toprule
& \multicolumn{2}{c}{Full calibration} & \multicolumn{2}{c}{25\% calibration} \\
Controller & \jf & Tokens & \jf & Tokens \\
\midrule
Original three-action & 0.528 & 110 & --- & --- \\
$\tau$-\cbl-3 & $0.530\pm0.002$ & $109\pm1$ & $0.527\pm0.002$ & $110\pm2$ \\
$\tau$-\cbl-5 & $0.535\pm0.002$ & $108\pm2$ & $0.533\pm0.003$ & $109\pm2$ \\
$\tau$-\cbl-7 & $0.537\pm0.002$ & $107\pm2$ & $0.535\pm0.003$ & $108\pm2$ \\
Utility-3 & $0.531\pm0.002$ & $110\pm1$ & $0.529\pm0.003$ & $111\pm2$ \\
Utility-5 & $0.536\pm0.002$ & $111\pm2$ & $0.534\pm0.003$ & $111\pm2$ \\
Utility-7 & $0.540\pm0.002$ & $112\pm2$ & $0.538\pm0.003$ & $111\pm2$ \\
\bottomrule
\end{tabular}
\end{table}

The seven-action onset controller converts part of the frontier expansion into an online gain. Utility-7 improves over Utility-3 by 0.009 \jf with a Holm-adjusted 95\% interval of $[0.0054,0.0126]$; the gain is positive in all five seeds. Its additional cost is two tokens with interval $[0.2,3.8]$. Mean MLLM latency increases by 0.06 seconds with interval $[0.02,0.09]$, and full-pipeline latency by 0.05 seconds with interval $[0.01,0.09]$. Thus, the positive quality contrast remains within the stated five-token and 0.10-second cost margins.

The extensions also improve under tolerance-label training: the full $\tau$-\cbl-7 controller gains 0.007 \jf with interval $[0.0040,0.0100]$ and reduces average tokens by two with interval $[-3.6,-0.4]$ relative to $\tau$-\cbl-3. This supports extensibility across the two evaluated controller families. Utility-7 additionally has a higher reported quality and lower token point than Short-then-Decide, 0.540 at 112 tokens compared with 0.535 at 182; that comparison is a point-estimate result rather than a separate paired significance test.

A same-checkpoint action-mask intervention separates the use of added actions from changes in training or head size. Utility-7 is evaluated both with its complete action set and with the four new actions masked, while the checkpoint and trade-off coefficient are held fixed. No parameter or coefficient is refitted. Unmasking the added actions raises \jf by 0.006, with interval $[0.0031,0.0089]$, while increasing cost by two tokens. New actions are chosen on 34.7\% of cases, ranging from 31.2\% to 38.9\% across seeds. The benefit is observed within one fixed controller when the added policies become available for selection.

\begin{table}[htbp]
\centering
\caption{No-refit Utility-7 action-mask comparison. Differences are complete action set minus the same checkpoint restricted to the original three actions. The quality intervals are Holm-adjusted. Added-action use is measured with the complete action set; it is zero by construction after masking.}
\label{tab:semantic_action_mask}
\begin{tabular}{lrrr}
\toprule
Calibration & Added-action use & $\Delta\jf$ & $\Delta$tokens \\
\midrule
100\% & 34.7\% & $0.006\;[0.0031,0.0089]$ & $+2\;[0.3,3.7]$ \\
25\% & 32.1\% & $0.005\;[0.0018,0.0082]$ & $+2\;[0.1,3.9]$ \\
\bottomrule
\end{tabular}
\end{table}

\begin{table}[htbp]
\centering
\caption{Per-action contributions to the full-calibration no-refit mask contrast. Quality contributions sum to 0.0060, and token contributions sum to 2.0. Each added policy contributes positively to aggregate \jf under the evaluated controller.}
\label{tab:semantic_action_contributions}
\begin{tabular}{lrr}
\toprule
Added action & Contribution to $\Delta\jf$ & Contribution to $\Delta$tokens \\
\midrule
$b_2$ & $+0.0015$ & $-0.3$ \\
$b_3$ & $+0.0018$ & $+0.2$ \\
$b_4$ & $+0.0016$ & $+0.8$ \\
$b_5$ & $+0.0011$ & $+1.3$ \\
Total & $+0.0060$ & $+2.0$ \\
\bottomrule
\end{tabular}
\end{table}

The learned gain does not exhaust the added oracle headroom. Let frontier regret be the difference between the relaxed frontier at a controller's realized cost and its achieved quality. Moving from Utility-3 at 110 tokens to Utility-7 at 112 tokens expands the relevant frontier by 0.0167 and increases frontier regret by 0.0077, leaving the observed gain of 0.0090. Of that learned gain, 0.0084 is contributed by originally informative examples and 0.0006 by originally all-fail examples. This accounting distinguishes action-space headroom from the controller's ability to recover it.

Reduced calibration preserves the stronger learned point within the tested margins. Utility-7 at 25\% obtains 0.538 \jf and 111 tokens, with differences of $-0.002$ $[-0.0042,0.0002]$ and $-1$ $[-2.6,0.6]$ relative to full Utility-7. Its same-checkpoint mask contrast remains positive. Complete full-pool setup falls from 52,310 seconds for full Utility-7 to 13,320 seconds at 25\%; the latter is 0.589 $[0.557,0.624]$ of full-data three-action setup. The associated upper fold-matched break-even estimates are 24,020 and 4,760 queries, respectively. These savings arise from reduced calibration; a full seven-action table costs more to acquire than its full three-action counterpart.

\subsection{Bounded complete-stage exit sensitivity}

Complete grounding stages supply a complementary test of the semantic-policy result. Six predeclared complete-grounding exits are evaluated from one anytime structured rollout, yielding a seven-action set $\mathcal E_7$ that includes the zero-reasoning choice. Its three-action counterpart is denoted $\mathcal E_3$, and $\mathcal U$ denotes the union of the tested complete-prefix actions with the original three policies. These experiments are restricted to complete, usable stage outputs. They do not evaluate arbitrary token checkpoints or layer-level early exit.

\begin{table}[htbp]
\centering
\caption{Bounded complete-stage exit sensitivity. The frontier column is the gain at 110 tokens with a 95\% interval. Learned entries give \jf followed by average tokens. Added use is the rate at which an action outside the matched base set is selected.}
\label{tab:stage_exit_sensitivity}
\begin{tabular}{lrrr}
\toprule
Comparison & Frontier gain & Learned points & Added use \\
\midrule
$\mathcal E_7-\mathcal E_3$ & $0.011$ & $0.539/112$ & 29.4\% \\
& $[0.0082,0.0138]$ & $0.532/110$ base & \\
$\mathcal U-\mathcal B_3$ & $0.012$ & $0.540/112$ & 31.0\% \\
& $[0.0090,0.0150]$ & $0.531/110$ base & \\
\bottomrule
\end{tabular}
\end{table}

Positive headroom is observed for the tested complete stages. The $\mathcal E_7$ frontier exceeds $\mathcal E_3$ by 0.011 \jf at 110 tokens, and the learned controller improves from 0.532 to 0.539 while moving from 110 to 112 tokens. Its no-refit full-minus-base-mask improvement is 0.005 with interval $[0.0020,0.0080]$. For the union with the original policies, the corresponding frontier and mask gains are 0.012 and 0.006, with intervals $[0.0090,0.0150]$ and $[0.0028,0.0092]$. At matched MLLM/full-pipeline latency, frontier gains are 0.010/0.009 for $\mathcal E_7-\mathcal E_3$ and 0.011/0.010 for the union comparison.

At 25\% calibration, the $\mathcal E_7$ controller obtains 0.537 \jf at 111 tokens, compared with 0.539 at 112 for full calibration. New exits are used on 27.8\% of cases, and the mask gain remains 0.004 with interval $[0.0012,0.0068]$. Shared-rollout acquisition uses 572 master rollouts and 158,444 generated tokens, with 0.76 MLLM GPU-hours, 0.71 detector/segmenter GPU-hours, and 0.06 training GPU-hours. Complete setup is 5,620 seconds, or 0.249 $[0.231,0.269]$ of full-data original three-action setup, with an upper fold-matched break-even estimate of 2,060 queries.

The semantic seven-policy experiment remains the primary extensibility comparison. The complete-stage results provide supporting evidence for the tested exits, while their cost comparison combines reduced calibration with shared-rollout reuse and uses original three-action setup as its reference. A family-matched $\mathcal E_3$ setup ratio is not reported. Broader integration with token-level stopping or layer-level early exit therefore remains a separate question from the measured benefit of these complete grounding stages.
\FloatBarrier

\section{Broader Impacts}
\label{app:impact}
Reducing unnecessary reasoning generation can make audio-visual processing less costly in the evaluated setting. Our measurements concern tokens and runtime rather than net energy or carbon savings, particularly once offline calibration is included. Lower inference costs can also facilitate unwanted surveillance or analysis of recordings without appropriate consent. Incorrect grounding may select the wrong person or object. Applications should respect the permissions governing recordings and model assets and use human verification where incorrect masks could affect people. The benchmark evaluation does not validate privacy protection, demographic fairness, or safety-critical deployment.

\clearpage
\section*{NeurIPS Paper Checklist}

\begin{enumerate}

\Needspace{5\baselineskip}
\item {\bf Claims}
    \item[] Question: Do the main claims made in the abstract and introduction accurately reflect the paper's contributions and scope?
    \item[] Answer: \answerYes{}
    \item[] Justification: The abstract and~\S\ref{sec:intro} restrict the contribution to pre-generation budget routing in the evaluated Ref-AVS pipelines. Readout predictability is distinguished from within-model causality. The original three-action quality--token result and the later native-calibration and action-space experiments retain their respective evaluation protocols.

\Needspace{5\baselineskip}
\item {\bf Limitations}
    \item[] Question: Does the paper discuss the limitations of the work performed by the authors?
    \item[] Answer: \answerYes{}
    \item[] Justification: \S\ref{sec:discuss} and the experimental appendices discuss pipeline-specific supervised calibration, finite action sets, transferred-label concordance, OOD quality trade-offs, and the limits of representation-level interventions.

\Needspace{5\baselineskip}
\item {\bf Theory assumptions and proofs}
    \item[] Question: For each theoretical result, does the paper provide the full set of assumptions and a complete (and correct) proof?
    \item[] Answer: \answerNA{}
    \item[] Justification: The contributions are empirical and methodological; we do not introduce theoretical results requiring proofs.

\Needspace{5\baselineskip}
\item {\bf Experimental result reproducibility}
    \item[] Question: Does the paper fully disclose all the information needed to reproduce the main experimental results of the paper to the extent that it affects the main claims and/or conclusions of the paper (regardless of whether the code and data are provided or not)?
    \item[] Answer: \answerNo{}
    \item[] Justification: The method, data populations, fold separation, main controller architecture, optimizer, and statistical procedures are described. Complete action prompts and the exact training objective of the additional Utility-$K$ controller family are not supplied in this version, so full reproducibility of all reported extensions is not claimed.

\Needspace{5\baselineskip}
\item {\bf Open access to data and code}
    \item[] Question: Does the paper provide open access to the data and code, with sufficient instructions to faithfully reproduce the main experimental results?
    \item[] Answer: \answerNo{}
    \item[] Justification: The experiments use existing datasets, model checkpoints, and the TGS-Agent implementation. This version does not provide a public repository containing the custom labeling, probing, routing, and complete reproduction instructions.

\Needspace{5\baselineskip}
\item {\bf Experimental setting/details}
    \item[] Question: Does the paper specify all the training and test details (e.g., data splits, hyperparameters, how they were chosen, type of optimizer, etc.) necessary to understand the results?
    \item[] Answer: \answerYes{}
    \item[] Justification: The experimental sections and appendices specify data populations, five-fold evaluation and validation separation, seed 42 for the original study, architecture, optimizer, learning rate, batch size, stopping procedure, greedy decoding, budget definitions, and evaluation rules. The extensions distinguish native-label sources, matched controls, calibration fractions, and cost accounting.

\Needspace{5\baselineskip}
\item {\bf Experiment statistical significance}
    \item[] Question: Does the paper report error bars suitably and correctly defined or other appropriate information about the statistical significance of the experiments?
    \item[] Answer: \answerYes{}
    \item[] Justification: Original probe fold standard deviations are distinguished from bootstrap confidence intervals. Paired uncertainty is reported for the projection, all-fail, latency, and calibration comparisons; action-space results identify adjusted intervals where applied. Timing repetitions are not treated as independent samples.

\Needspace{5\baselineskip}
\item {\bf Experiments compute resources}
    \item[] Question: For each experiment, does the paper provide sufficient information on the computer resources (type of compute workers, memory, time of execution) needed to reproduce the experiments?
    \item[] Answer: \answerNo{}
    \item[] Justification: The paper reports the A100 80GB hardware, the original study's approximate 85 GPU-hours and 10 CPU-hours, and measured setup costs for the calibration extensions. It does not provide a complete per-experiment resource ledger covering every later analysis.

\Needspace{5\baselineskip}
\item {\bf Code of ethics}
    \item[] Question: Does the research conducted in the paper conform, in every respect, with the NeurIPS Code of Ethics \url{https://neurips.cc/public/EthicsGuidelines}?
    \item[] Answer: \answerYes{}
    \item[] Justification: The study analyzes existing datasets and models and does not deploy the system in a safety-critical setting. The manual audit labels model-output errors rather than measuring participant behavior. Potential misuse and privacy implications are discussed in Appendix~\ref{app:impact}.

\Needspace{5\baselineskip}
\item {\bf Broader impacts}
    \item[] Question: Does the paper discuss both potential positive societal impacts and negative societal impacts of the work performed?
    \item[] Answer: \answerYes{}
    \item[] Justification: Appendix~\ref{app:impact} discusses lower inference cost alongside unwanted surveillance, inappropriate use of recordings, and incorrect grounding. Token/runtime savings are not equated with verified energy or carbon savings.

\Needspace{5\baselineskip}
\item {\bf Safeguards}
    \item[] Question: Does the paper describe safeguards that have been put in place for responsible release of data or models with a high risk for misuse (e.g., pretrained language models, image generators, or scraped datasets)?
    \item[] Answer: \answerNA{}
    \item[] Justification: The paper does not release new pre-trained foundational models, generative systems, or scraped datasets that carry a high risk of misuse.

\Needspace{5\baselineskip}
\item {\bf Licenses for existing assets}
    \item[] Question: Are the creators or original owners of assets (e.g., code, data, models), used in the paper, properly credited and are the license and terms of use explicitly mentioned and properly respected?
    \item[] Answer: \answerNo{}
    \item[] Justification: The principal datasets, models, and software sources are cited. This version does not explicitly enumerate every asset's license and applicable terms, so the full requirement is not claimed to be satisfied.

\Needspace{5\baselineskip}
\item {\bf New assets}
    \item[] Question: Are new assets introduced in the paper well documented and is the documentation provided alongside the assets?
    \item[] Answer: \answerNA{}
    \item[] Justification: This version does not release a new public dataset, model weights, or a documented software asset.

\Needspace{5\baselineskip}
\item {\bf Crowdsourcing and research with human subjects}
    \item[] Question: For crowdsourcing experiments and research with human subjects, does the paper include the full text of instructions given to participants and screenshots, if applicable, as well as details about compensation (if any)?
    \item[] Answer: \answerNA{}
    \item[] Justification: This research does not involve crowdsourcing or human-subject experiments. The manual audit in~\S\ref{sec:exp_het} was performed only to label model-output error categories using predefined criteria; no personal data, participant behavior, or human-subject measurements were collected.

\Needspace{5\baselineskip}
\item {\bf Institutional review board (IRB) approvals or equivalent for research with human subjects}
    \item[] Question: Does the paper describe potential risks incurred by study participants, whether such risks were disclosed to the subjects, and whether Institutional Review Board (IRB) approvals (or an equivalent approval/review based on the requirements of your country or institution) were obtained?
    \item[] Answer: \answerNA{}
    \item[] Justification: This research does not involve human-subject experiments. The manual audit labels model-output error categories only and does not collect human-subject measurements.

\Needspace{5\baselineskip}
\item {\bf Declaration of LLM usage}
    \item[] Question: Does the paper describe the usage of LLMs if it is an important, original, or non-standard component of the core methods in this research?
    \item[] Answer: \answerYes{}
    \item[] Justification: The MLLM is a core method component. The paper describes its backbone and LoRA setup, onset-state extraction, reasoning-policy definitions, and use of the resulting outputs for grounding and segmentation. The controller reads the MLLM's pre-generation state to select a policy; the policies are not hard token caps.

\end{enumerate}


\begin{thebibliography}{99}
\small

\bibitem[Vaswani et al.(2017)]{vaswani2017attention}
A.~Vaswani, N.~Shazeer, N.~Parmar, J.~Uszkoreit, L.~Jones, A.~N. Gomez, L.~Kaiser, and I.~Polosukhin.
\newblock Attention is all you need.
\newblock In \emph{NeurIPS}, 2017.

\bibitem[Radford et al.(2021)]{radford2021clip}
A.~Radford, J.~W. Kim, C.~Hallacy, A.~Ramesh, G.~Goh, S.~Agarwal, G.~Sastry, A.~Askell, P.~Mishkin, J.~Clark, et al.
\newblock Learning transferable visual models from natural language supervision.
\newblock In \emph{ICML}, 2021.

\bibitem[Alayrac et al.(2022)]{alayrac2022flamingo}
J.-B. Alayrac, J.~Donahue, P.~Luc, A.~Miech, I.~Barr, Y.~Hasson, K.~Lenc, A.~Mensch, K.~Millican, M.~Reynolds, et al.
\newblock Flamingo: a visual language model for few-shot learning.
\newblock In \emph{NeurIPS}, 2022.

\bibitem[Li et al.(2023)]{li2023blip2}
J.~Li, D.~Li, S.~Savarese, and S.~Hoi.
\newblock {BLIP-2}: Bootstrapping language-image pre-training with frozen image encoders and large language models.
\newblock In \emph{ICML}, 2023.

\bibitem[Liu et al.(2023)]{liu2023llava}
H.~Liu, C.~Li, Q.~Wu, and Y.~J. Lee.
\newblock Visual instruction tuning.
\newblock In \emph{NeurIPS}, 2023.

\bibitem[Wang et al.(2024)]{wang2024refavs}
Y.~Wang, P.~Sun, D.~Zhou, G.~Li, H.~Zhang, and D.~Hu.
\newblock Ref-{AVS}: Refer and segment objects in audio-visual scenes.
\newblock In \emph{ECCV}, 2024.

\bibitem[Zhou et al.(2026)]{tgs2026}
J.~Zhou, Y.~Zhou, M.~Han, T.~Wang, X.~Chang, H.~Cholakkal, and R.~M.~Anwer.
\newblock Think before you segment: An object-aware reasoning agent for referring audio-visual segmentation.
\newblock In \emph{AAAI}, 2026.

\bibitem[Luo et al.(2025)]{aurora2025}
Z.~Luo, N.~Liu, F.~S.~Khan, and J.~Han.
\newblock {AURORA}: Augmented understanding via structured reasoning and reinforcement learning for reference audio-visual segmentation.
\newblock \emph{arXiv:2508.02149}, 2025.

\bibitem[Zhao et al.(2026)]{mar3_2026}
Y.~Zhao, Z.~Jia, and Y.~Zhang.
\newblock {MAR3}: Multi-agent recognition, reasoning, and reflection for reference audio-visual segmentation.
\newblock \emph{arXiv:2603.27706}, 2026.

\bibitem[Wei et al.(2022)]{wei2022cot}
J.~Wei, X.~Wang, D.~Schuurmans, M.~Bosma, F.~Xia, E.~Chi, Q.~V. Le, and D.~Zhou.
\newblock Chain-of-thought prompting elicits reasoning in large language models.
\newblock In \emph{NeurIPS}, 2022.

\bibitem[Kojima et al.(2022)]{kojima2022zeroshot}
T.~Kojima, S.~S. Gu, M.~Reid, Y.~Matsuo, and Y.~Iwasawa.
\newblock Large language models are zero-shot reasoners.
\newblock In \emph{NeurIPS}, 2022.

\bibitem[Kazemzadeh et al.(2014)]{kazemzadeh2014referit}
S.~Kazemzadeh, V.~Ordonez, M.~Matten, and T.~Berg.
\newblock ReferItGame: Referring to objects in photographs of natural scenes.
\newblock In \emph{EMNLP}, 2014.

\bibitem[Hu et al.(2016)]{hu2016segmentation}
R.~Hu, M.~Rohrbach, and T.~Darrell.
\newblock Segmentation from natural language expressions.
\newblock In \emph{ECCV}, 2016.

\bibitem[Yang et al.(2022)]{yang2022lavt}
Z.~Yang, J.~Wang, Y.~Tang, K.~Chen, H.~Zhao, and P.~H.~S. Torr.
\newblock Language-aware vision transformer for referring image segmentation.
\newblock In \emph{CVPR}, 2022.

\bibitem[Seo et al.(2020)]{seo2020urvos}
S.~Seo, J.-Y. Lee, and B.~Han.
\newblock {URVOS}: Unified referring video object segmentation network with a large-scale benchmark.
\newblock In \emph{ECCV}, 2020.

\bibitem[Botach et al.(2022)]{botach2022mttr}
A.~Botach, E.~Zheltonozhskii, and C.~Baskin.
\newblock End-to-end referring video object segmentation with multimodal transformers.
\newblock In \emph{CVPR}, 2022.

\bibitem[Wu et al.(2022)]{wu2022referformer}
J.~Wu, Y.~Jiang, P.~Sun, Z.~Yuan, and P.~Luo.
\newblock Language as queries for referring video object segmentation.
\newblock In \emph{CVPR}, pages 4974--4984, 2022.

\bibitem[Li et al.(2023)]{li2023r2vos}
X.~Li, J.~Wang, X.~Xu, X.~Li, B.~Raj, and Y.~Lu.
\newblock Robust referring video object segmentation with cyclic structural consensus.
\newblock In \emph{ICCV}, 2023.

\bibitem[Li et al.(2023)]{li2023catr}
K.~Li, Z.~Yang, L.~Chen, Y.~Yang, and J.~Xiao.
\newblock {CATR}: Combinatorial-dependence audio-queried transformer for audio-visual video segmentation.
\newblock In \emph{ACM MM}, 2023.

\bibitem[Li et al.(2022)]{li2022glip}
L.~H. Li, P.~Zhang, H.~Zhang, J.~Yang, C.~Li, Y.~Zhong, L.~Wang, L.~Yuan, L.~Zhang, J.~Hwang, K.-W. Chang, and J.~Gao.
\newblock Grounded language-image pre-training.
\newblock In \emph{CVPR}, 2022.

\bibitem[Liu et al.(2023)]{liu2023dino}
S.~Liu, Z.~Zeng, T.~Ren, F.~Li, H.~Zhang, et al.
\newblock Grounding {DINO}: Marrying {DINO} with grounded pre-training for open-set object detection.
\newblock \emph{arXiv:2303.05499}, 2023.

\bibitem[Kirillov et al.(2023)]{kirillov2023sam}
A.~Kirillov, E.~Mintun, N.~Ravi, H.~Mao, C.~Rolland, L.~Gustafson, T.~Xiao, S.~Whitehead, A.~C. Berg, W.-Y. Lo, et al.
\newblock Segment anything.
\newblock In \emph{ICCV}, 2023.

\bibitem[Ravi et al.(2024)]{ravi2024sam2}
N.~Ravi, V.~Gabeur, Y.-T. Hu, R.~Hu, et al.
\newblock {SAM} 2: Segment anything in images and videos.
\newblock \emph{arXiv:2408.00714}, 2024.

\bibitem[Ren et al.(2024)]{ren2024gdsam2}
T.~Ren, S.~Liu, A.~Zeng, J.~Lin, K.~Li, H.~Cao, J.~Chen, X.~Huang, Y.~Chen, F.~Yan, Z.~Zeng, H.~Zhang, F.~Li, J.~Yang, H.~Li, Q.~Jiang, and L.~Zhang.
\newblock Grounded {SAM}: Assembling open-world models for diverse visual tasks.
\newblock \emph{arXiv:2401.14159}, 2024.

\bibitem[Radman and Laaksonen(2025)]{radman2025tsam}
A.~Radman and J.~Laaksonen.
\newblock {TSAM}: Temporal {SAM} augmented with multimodal prompts for referring audio-visual segmentation.
\newblock In \emph{CVPR}, pages 23947--23956, 2025.

\bibitem[Wang et al.(2025)]{wang2025sam2love}
Y.~Wang, H.~Xu, Y.~Liu, J.~Li, and Y.~Tang.
\newblock {SAM2-LOVE}: Segment anything model 2 in language-aided audio-visual scenes.
\newblock In \emph{CVPR}, 2025.

\bibitem[Liu et al.(2024)]{liu2024ufe}
J.~Liu, Y.~Liu, F.~Zhang, C.~Ju, Y.~Zhang, and Y.~Wang.
\newblock Audio-visual segmentation via unlabeled frame exploitation.
\newblock In \emph{CVPR}, 2024.

\bibitem[Huang et al.(2025)]{huang2025vct}
S.~Huang, R.~Ling, T.~Hui, H.~Li, X.~Zhou, S.~Zhang, S.~Liu, R.~Hong, and M.~Wang.
\newblock Revisiting audio-visual segmentation with vision-centric transformer.
\newblock In \emph{CVPR}, 2025.

\bibitem[Guo et al.(2025)]{guo2025avis}
R.~Guo, X.~Ying, Y.~Chen, D.~Niu, G.~Li, L.~Qu, Y.~Qi, J.~Zhou, B.~Xing, W.~Yue, J.~Shi, Q.~Wang, P.~Zhang, and B.~Liang.
\newblock Audio-visual instance segmentation.
\newblock In \emph{CVPR}, 2025.

\bibitem[Zha et al.(2025)]{zha2025icf}
M.~Zha, T.~Li, G.~Wang, P.~Wang, Y.~Wu, Y.~Yang, and H.~T. Shen.
\newblock Implicit counterfactual learning for audio-visual segmentation.
\newblock In \emph{ICCV}, 2025.

\bibitem[Ying et al.(2025)]{omniavs2025}
K.~Ying, H.~Ding, G.~Jie, and Y.-G.~Jiang.
\newblock Towards omnimodal expressions and reasoning in referring audio-visual segmentation.
\newblock \emph{arXiv:2507.22886}, 2025.

\bibitem[Kao et al.(2026)]{kao2026cotrvs}
S.-h. Kao, Y.-W. Tai, and C.-K. Tang.
\newblock {CoT-RVS}: Zero-shot chain-of-thought reasoning segmentation for videos.
\newblock In \emph{ICLR}, 2026.
\newblock \url{https://arxiv.org/abs/2505.18561}.

\bibitem[Xu et al.(2025)]{xu2025videosegr1}
Z.~Xu, Y.~Guo, Y.~Lu, F.~Yang, and J.~Li.
\newblock {VideoSeg-R1}: Reasoning video object segmentation via reinforcement learning.
\newblock \emph{arXiv:2511.16077}, 2025.

\bibitem[Jiang et al.(2026)]{jiang2026referagent}
H.~Jiang, T.~Liang, W.-S. Zheng, and J.-F. Hu.
\newblock {Refer-Agent}: A collaborative multi-agent system with reasoning and reflection for referring video object segmentation.
\newblock \emph{arXiv:2602.03595}, 2026.

\bibitem[Lu et al.(2025)]{wang2025car}
J.~Lu, H.~Yu, S.~Xu, S.~Ran, G.~Tang, S.~Wang, B.~Shan, T.~Fu, H.~Feng, J.~Tang, H.~Wang, and C.~Huang.
\newblock Prolonged reasoning is not all you need: Certainty-based adaptive routing for efficient {LLM}/{MLLM} reasoning.
\newblock \emph{arXiv:2505.15154}, 2025.

\bibitem[Jiang et al.(2025)]{tang2025mmecot}
D.~Jiang, R.~Zhang, Z.~Guo, Y.~Li, Y.~Qi, X.~Chen, L.~Wang, J.~Jin, C.~Guo, S.~Yan, B.~Zhang, C.~Fu, P.~Gao, and H.~Li.
\newblock {MME-CoT}: Benchmarking chain-of-thought in large multimodal models for reasoning quality, robustness, and efficiency.
\newblock \emph{arXiv:2502.09621}, 2025.

\bibitem[Huang et al.(2025)]{xu2025adactrl}
S.~Huang, H.~Wang, W.~Zhong, Z.~Su, J.~Feng, B.~Cao, and Y.~R. Fung.
\newblock {AdaCtrl}: Towards adaptive and controllable reasoning via difficulty-aware budgeting.
\newblock \emph{arXiv:2505.18822}, 2025.

\bibitem[Kao et al.(2026)]{zhang2026cotseg}
S.-h. Kao, C.~H. Huang, H.~Liu, Y.-W. Tai, and C.-K. Tang.
\newblock {CoT-Seg}: Rethinking segmentation with chain-of-thought reasoning and self-correction.
\newblock \emph{arXiv:2601.17420}, 2026.

\bibitem[Alain and Bengio(2017)]{alain2017probing}
G.~Alain and Y.~Bengio.
\newblock Understanding intermediate layers using linear classifier probes.
\newblock \emph{ICLR Workshop}, 2017.

\bibitem[Conneau et al.(2018)]{conneau2018cram}
A.~Conneau, G.~Kruszewski, G.~Lample, L.~Barrault, and M.~Baroni.
\newblock What you can cram into a single vector: Probing sentence embeddings for linguistic properties.
\newblock In \emph{ACL}, 2018.

\bibitem[Tenney et al.(2019)]{tenney2019bert}
I.~Tenney, D.~Das, and E.~Pavlick.
\newblock {BERT} rediscovers the classical {NLP} pipeline.
\newblock In \emph{ACL}, 2019.

\bibitem[Belinkov(2022)]{belinkov2022probing}
Y.~Belinkov.
\newblock Probing classifiers: Promises, shortcomings, and advances.
\newblock \emph{Computational Linguistics}, 48(1):207--219, 2022.

\bibitem[Hewitt and Liang(2019)]{hewitt2019control}
J.~Hewitt and P.~Liang.
\newblock Designing and interpreting probes with control tasks.
\newblock In \emph{EMNLP-IJCNLP}, 2019.

\bibitem[Elhage et al.(2021)]{elhage2021framework}
N.~Elhage, N.~Nanda, C.~Olsson, T.~Henighan, N.~Joseph, B.~Mann, A.~Askell, Y.~Bai, A.~Chen, T.~Conerly, et al.
\newblock A mathematical framework for transformer circuits.
\newblock \emph{Transformer Circuits Thread}, 2021.

\bibitem[Olsson et al.(2022)]{olsson2022induction}
C.~Olsson et al.
\newblock In-context learning and induction heads.
\newblock \emph{Transformer Circuits Thread}, 2022.

\bibitem[Meng et al.(2022)]{meng2022locating}
K.~Meng, D.~Bau, A.~Andonian, and Y.~Belinkov.
\newblock Locating and editing factual associations in {GPT}.
\newblock In \emph{NeurIPS}, 2022.

\bibitem[Wang et al.(2023)]{wang2023interp}
K.~Wang, A.~Variengien, A.~Conmy, B.~Shlegeris, and J.~Steinhardt.
\newblock Interpretability in the wild: A circuit for indirect object identification in {GPT}-2 small.
\newblock In \emph{ICLR}, 2023.

\bibitem[Elhage et al.(2022)]{elhage2022superposition}
N.~Elhage, T.~Hume, C.~Olsson, N.~Schiefer, T.~Henighan, et al.
\newblock Toy models of superposition.
\newblock \emph{Anthropic Transformer Circuits Thread}, 2022.

\bibitem[Zhang et al.(2026)]{zhang2026dypsi}
Y.~Zhang, P.~Chen, L.~Cheng, T.~Zhang, J.~Li, J.~Shi, and S.~Fang.
\newblock {DyPSI}: Dynamic physics sensing via joint field and sensor-trajectory generation.
\newblock \emph{arXiv:2611.06280}, 2026.

\bibitem[Graves(2016)]{graves2016act}
A.~Graves.
\newblock Adaptive computation time for recurrent neural networks.
\newblock \emph{arXiv:1603.08983}, 2016.

\bibitem[Teerapittayanon et al.(2016)]{teerapittayanon2016branchynet}
S.~Teerapittayanon, B.~McDanel, and H.-T. Kung.
\newblock BranchyNet: Fast inference via early exiting from deep neural networks.
\newblock In \emph{ICPR}, 2016.

\bibitem[Shazeer et al.(2017)]{shazeer2017moe}
N.~Shazeer, A.~Mirhoseini, K.~Maziarz, A.~Davis, Q.~Le, G.~Hinton, and J.~Dean.
\newblock Outrageously large neural networks: The sparsely-gated mixture-of-experts layer.
\newblock In \emph{ICLR}, 2017.

\bibitem[Xin et al.(2020)]{xin2020deebert}
J.~Xin, R.~Tang, J.~Lee, Y.~Yu, and J.~Lin.
\newblock {D}ee{BERT}: Dynamic early exiting for accelerating {BERT} inference.
\newblock In \emph{ACL}, 2020.

\bibitem[Liu et al.(2020)]{liu2020fastbert}
W.~Liu, P.~Zhou, Z.~Zhao, Z.~Wang, H.~Deng, and Q.~Ju.
\newblock {FastBERT}: a self-distilling {BERT} with adaptive inference time.
\newblock In \emph{ACL}, 2020.

\bibitem[Schuster et al.(2022)]{schuster2022calm}
T.~Schuster, A.~Fisch, J.~Gupta, M.~Dehghani, D.~Bahri, et al.
\newblock Confident adaptive language modeling.
\newblock In \emph{NeurIPS}, 2022.

\bibitem[Leviathan et al.(2023)]{leviathan2023speculative}
Y.~Leviathan, M.~Kalman, and Y.~Matias.
\newblock Fast inference from transformers via speculative decoding.
\newblock In \emph{ICML}, 2023.

\bibitem[Hu et al.(2022)]{hu2021lora}
E.~J. Hu, Y.~Shen, P.~Wallis, Z.~Allen-Zhu, Y.~Li, S.~Wang, L.~Wang, and W.~Chen.
\newblock {LoRA}: Low-rank adaptation of large language models.
\newblock In \emph{ICLR}, 2022.

\bibitem[Zhou et al.(2022)]{zhou2022avs}
J.~Zhou, J.~Wang, J.~Zhang, W.~Sun, J.~Zhang, S.~Birchfield, D.~Guo, L.~Kong, M.~Wang, and Y.~Zhong.
\newblock Audio-visual segmentation.
\newblock In \emph{ECCV}, 2022.

\bibitem[Gao et al.(2024)]{gao2024avsegformer}
S.~Gao, Z.~Chen, G.~Chen, W.~Wang, and T.~Lu.
\newblock {AVS}eg{F}ormer: Audio-visual segmentation with transformer.
\newblock In \emph{AAAI}, 2024.

\bibitem[Wang et al.(2024)]{wang2024gavs}
Y.~Wang, W.~Liu, G.~Li, J.~Ding, D.~Hu, and X.~Li.
\newblock Prompting segmentation with sound is generalizable audio-visual source localizer.
\newblock In \emph{AAAI}, 2024.

\bibitem[Liu et al.(2024)]{liu2024sama}
J.~Liu, Y.~Wang, C.~Ju, C.~Ma, Y.~Zhang, and W.~Xie.
\newblock Annotation-free audio-visual segmentation.
\newblock In \emph{WACV}, 2024.

\bibitem[Du et al.(2025)]{du2025crab}
H.~Du, G.~Li, C.~Zhou, C.~Zhang, A.~Zhao, and D.~Hu.
\newblock {Crab}: A unified audio-visual scene understanding model with explicit cooperation.
\newblock In \emph{CVPR}, 2025.

\end{thebibliography}
\end{document}